\documentclass[aps,showpacs]{revtex4}
\usepackage{graphicx}
\usepackage{amssymb}

\begin{document}

\title{
Log-periodic behavior of finite size effects in
field theories with RG limit cycles
}
\author{Andr\'e  LeClair$^{\clubsuit , \spadesuit}$, 
Jos\'e Mar\'{\i}a Rom\'an$^\spadesuit$ and  Germ\'an Sierra$^\spadesuit$}
\affiliation{$^\clubsuit$Newman Laboratory, Cornell University, Ithaca, NY}
\affiliation{$^\spadesuit$Instituto de F\'{\i}sica Te\'orica, UAM-CSIC, Madrid, Spain} 
\date{November, 2003}

\begin{abstract}

We compute the finite size effects in the ground state energy,
equivalently the effective central charge $c_{\rm eff}$,  based on S-matrix
theories recently conjectured to describe a cyclic regime of 
the Kosterlitz-Thouless renormalization group flows.   
The effective central charge has periodic properties consistent  with 
renormalization group predictions.   Whereas  $c_{\rm eff}$ for 
 the massive case
has a singularity in the very deep ultra-violet,  we argue that the
massless 
version is non-singular and periodic on all length scales.

\end{abstract}


\maketitle

\vskip 0.2cm

%
%
%
%
\def\oti{{\otimes}}
\def\lb{ \left[ }
\def\rb{ \right]  }
\def\tilde{\widetilde}
\def\bar{\overline}
\def\hat{\widehat}
\def\*{\star}
\def\[{\left[}
\def\]{\right]}
\def\({\left(}		\def\BL{\Bigr(}
\def\){\right)}		\def\BR{\Bigr)}
	\def\BBL{\lb}
	\def\BBR{\rb}
%
%
\def\zb{{\bar{z} }}
\def\zbar{{\bar{z} }}
\def\frac#1#2{{#1 \over #2}}
\def\inv#1{{1 \over #1}}
\def\half{{1 \over 2}}
\def\d{\partial}
\def\der#1{{\partial \over \partial #1}}
\def\dd#1#2{{\partial #1 \over \partial #2}}
\def\vev#1{\langle #1 \rangle}
\def\ket#1{ | #1 \rangle}
\def\rvac{\hbox{$\vert 0\rangle$}}
\def\lvac{\hbox{$\langle 0 \vert $}}
\def\2pi{\hbox{$2\pi i$}}
\def\e#1{{\rm e}^{^{\textstyle #1}}}
\def\grad#1{\,\nabla\!_{{#1}}\,}
\def\dsl{\raise.15ex\hbox{/}\kern-.57em\partial}
\def\Dsl{\,\raise.15ex\hbox{/}\mkern-.13.5mu D}
%
%
\def\ga{\gamma}		\def\Ga{\Gamma}
\def\be{\beta}
\def\al{\alpha}
\def\ep{\epsilon}
\def\vep{\varepsilon}
\def\la{\lambda}	\def\La{\Lambda}
\def\de{\delta}		\def\De{\Delta}
\def\om{\omega}		\def\Om{\Omega}
\def\sig{\sigma}	\def\Sig{\Sigma}
\def\vphi{\varphi}
%
%
\def\CA{{\cal A}}	\def\CB{{\cal B}}	\def\CC{{\cal C}}
\def\CD{{\cal D}}	\def\CE{{\cal E}}	\def\CF{{\cal F}}
\def\CG{{\cal G}}	\def\CH{{\cal H}}	\def\CI{{\cal J}}
\def\CJ{{\cal J}}	\def\CK{{\cal K}}	\def\CL{{\cal L}}
\def\CM{{\cal M}}	\def\CN{{\cal N}}	\def\CO{{\cal O}}
\def\CP{{\cal P}}	\def\CQ{{\cal Q}}	\def\CR{{\cal R}}
\def\CS{{\cal S}}	\def\CT{{\cal T}}	\def\CU{{\cal U}}
\def\CV{{\cal V}}	\def\CW{{\cal W}}	\def\CX{{\cal X}}
\def\CY{{\cal Y}}	\def\CZ{{\cal Z}}

\def\rvac{\hbox{$\vert 0\rangle$}}
\def\lvac{\hbox{$\langle 0 \vert $}}
\def\comm#1#2{ \BBL\ #1\ ,\ #2 \BBR }
\def\2pi{\hbox{$2\pi i$}}
\def\e#1{{\rm e}^{^{\textstyle #1}}}
\def\grad#1{\,\nabla\!_{{#1}}\,}
\def\dsl{\raise.15ex\hbox{/}\kern-.57em\partial}
\def\Dsl{\,\raise.15ex\hbox{/}\mkern-.13.5mu D}
%
%
%
\font\numbers=cmss12
\font\upright=cmu10 scaled\magstep1
\def\stroke{\vrule height8pt width0.4pt depth-0.1pt}
\def\topfleck{\vrule height8pt width0.5pt depth-5.9pt}
\def\botfleck{\vrule height2pt width0.5pt depth0.1pt}
\def\Zmath{\vcenter{\hbox{\numbers\rlap{\rlap{Z}\kern
0.8pt\topfleck}\kern 2.2pt
                   \rlap Z\kern 6pt\botfleck\kern 1pt}}}
\def\Qmath{\vcenter{\hbox{\upright\rlap{\rlap{Q}\kern
                   3.8pt\stroke}\phantom{Q}}}}
\def\Nmath{\vcenter{\hbox{\upright\rlap{I}\kern 1.7pt N}}}
\def\Cmath{\vcenter{\hbox{\upright\rlap{\rlap{C}\kern
                   3.8pt\stroke}\phantom{C}}}}
\def\Rmath{\vcenter{\hbox{\upright\rlap{I}\kern 1.7pt R}}}
\def\Z{\ifmmode\Zmath\else$\Zmath$\fi}
\def\Q{\ifmmode\Qmath\else$\Qmath$\fi}
\def\N{\ifmmode\Nmath\else$\Nmath$\fi}
\def\C{\ifmmode\Cmath\else$\Cmath$\fi}
\def\R{\ifmmode\Rmath\else$\Rmath$\fi}

\def\barray{\begin{eqnarray}}
\def\earray{\end{eqnarray}}
\def\beq{\begin{equation}}
\def\eeq{\end{equation}}

\def\no{\noindent}

\def\gpar{g_\parallel}
\def\gperp{g_\perp}

\def\Jb{\bar{J}}
\def\dx{\frac{d^2 x}{2\pi}}

\def\rap{\beta}
\def\s{\sigma}
\def\spec{\zeta}
\def\comb{\frac{\rap\theta}{2\pi} }
\def\Ga{\Gamma}
\def\eta{\alpha}

\def\L{{\cal L}}
\def\g{{\bf g}}
\def\K{{\cal K}}
\def\I{{\cal I}}
\def\M{{\cal M}}
\def\F{{\cal F}}

\def\gpar{g_\parallel}
\def\gperp{g_\perp}
\def\Jb{\bar{J}}
\def\dx{\frac{d^2 x}{2\pi}}
\def\imag{\Im {\it m}}
\def\real{\Re {\it e}}

\section{Introduction}

Over the last few years a novel renormalization group (RG) scenario
with limit-cycle behavior  
has been discovered in various models in a variety of physical
contexts, including nuclear physics \cite{nuclear},  
quantum field theory \cite{BLflow,LRS}, quantum mechanics \cite{GW},
superconductivity \cite{RDBCS}, Bose-Einstein condensation \cite{Bose},
and effective low energy  QCD \cite{QCD}.  
The subject of duality cascades in supersymmetric gauge theory \cite{Klebanov}
is also suggestive of limit-cycle behavior. 
The possibility of limit cycle behavior in the RG flow was
considered as early as 1971 by Wilson \cite{Kwilson}, however
at the time no models with this behavior were known.
The possibility of chaotic flows has also been recently
considered \cite{GW,morozov}.

As is readily realized,  models with cyclic RG flows have
novel physical properties in comparison with theories with
fixed points.   For theories with bound states,  the general
signature of the cyclic RG is an infinite tower of bound
states with energies related by discrete scaling relations \cite{nuclear,
GW,RDBCS,QCD}, i.e. a self-similarity of the spectrum under
a discrete scale transformation.   This property was termed
Russian doll scaling in  \cite{RDBCS} for obvious reasons. 
On the other hand,  for continuous quantum field
theories without bound states,  the cyclic RG is manifested in
the periodicity of the S-matrix as a function of energy \cite{LRS},
as anticipated already by Wilson \cite{Kwilson}.

\def\ceff{c_{\rm eff}}

The present article is a continuation of  \cite{LRS}.  In the latter
paper, an exact S-matrix for the cyclic regime of the 
Kosterlitz-Thouless flows was proposed for the first time.   Using
the S-matrices proposed there, here  we investigate the finite-size
effects, or equivalently the finite temperature effects,  
of the ground state energy $E(R)$   on a circle of circumference $R$
(temperature $1/R$).   Conventionally one defines the effective
central charge 
$c_{\rm eff}$ as $E(R) = - \pi \ceff (R)/6R$.  The quantity
$\ceff$ governs many physical properties, including the specific
heat \cite{Cardy,Affleck}.      It is known that for theories
with fixed points,  the quantity $\ceff$ tracks the RG flow,
namely, as a function of $R$ it smoothly interpolates between
the ultra-violet (UV),  $R=0$,   and infra-red (IR), $R=\infty$,
values of the Virasoro central
charge of the fixed points \cite{ZamoTBA}.   For this reason
$\ceff$ is an interesting property to study for a model with
an RG limit cycle,  and this is the main subject of this paper. 
As we will show,  it has a periodic structure as a function of
$\log R$.  

There is another c-function $c_z (L)$  which also tracks the RG flow 
between fixed points, this being the content of the celebrated
c-theorem \cite{Zamoc,Friedan}, which is a statement of the irreversibility
of the RG flow.  Specifically the theorem states that $c_z (L)$ 
is a monotonically decreasing function of scale $L$  in the flow toward
the IR.   This by itself seems to rule out cyclic RG flows in unitary
theories, since
straightforward RG arguments show that $c_z (L) $ should be periodic 
in $\log L$.  In fact the proof of the theorem constructs
$c_z$ as a function of the couplings $g$, and the flow of $c_z$ 
is induced from the flow of $g$: 
$dc_z  / d \log L =  \beta_g \d_g c_z (g)$,  where
$\beta_g $ is the $\beta$ function;  thus if the couplings
are periodic, so is $c_z$.     
Whereas $\ceff (R)$ is related to
a one-point function of the trace of the stress-energy tensor
in a finite geometry
(see below),  $c_z(L)$  is related to
a two-point function with fields separated by $L$ in  infinite volume, 
and they are thus
different.  Nevertheless, for models with fixed points 
both $\ceff (R)$ and $c_z (L)$ have essentially the same behavior,
and we expect the same to be true for a theory with a cyclic RG. 
This leaves us with a paradox, since our theory is a perturbation
of a unitary conformal field theory by a hermitian operator.
This issue will not be dealt with any further in this work.

Let us now summarize our main results.  In section II we outline
the main features one expects for integrable, 
 relativistic quantum field theories
in 2d that have a cyclic RG, using only standard RG arguments in
a model-independent way.   There a clear distinction is made between
massive and massless scattering theories.   In section III we
review the definition of the models and RG flows,  extending
the arguments to higher level $k$ current algebras.  We
point out two possible limiting theories,  one massive,  the other
massless,  depending on whether one has a UV or IR fixed point
in the isotropic limit.  In section IV we describe the proposed
S-matrices.  The massive S-matrix is the same as considered  in \cite{LRS},
whereas the massless one is new.    Both these S-matrices are periodic
in rapidity, which as explained in section II,  is a clear signature
of the cyclic RG.   In section V we use thermodynamic Bethe ansatz
techniques to study the finite size effects for the massive case. 
Though periodic structures consistent with the RG flows are 
found for $\ceff$, in the deep UV it develops a singularity
whose nature is explored in some detail.  
 We derive an approximate  analytic expression for 
$\ceff$ in terms of Riemann's zeta function,
 eq. (\ref{clinear}) which is surprisingly simple
and agrees very well with exact numerical results.   
 The finite size-effects
for the massless case are discussed in section VI; however at the
present time the appropriate thermodynamic Bethe ansatz equations
are unknown, and our analysis is thus incomplete.   However 
on general grounds we argue that in the massless case  $\ceff$  
should be  an exactly periodic function of $\log R$ on all scales and has
no singularities,  and thus appears  to be more consistent with
the RG analysis.

\section{Physical Consequences of an RG limit cycle}

In this section we describe some general properties 
that a model with an RG limit cycle should exhibit, as a
way of anticipating our subsequent results.  

Generally speaking, cyclic behavior  in the RG flow can in 
principle exist at all scales,  or can be approached  asymptotically
in the ultra-violet  or infra-red, the latter being
UV or IR limit-cycle behavior.  
  Once the flow
is in the cyclic regime, the couplings are periodic
\beq
\label{II.1}
g(l + \lambda) = g(l) 
\eeq
where $l= \log L$ and $L$ is the length scale.  Above,
the period $\lambda$ is fixed and model-dependent.  In our
model $\lambda$ is an RG invariant function of the two
running couplings.  As we will review in the next section,
our models have a periodic RG on all scales rather than
a limit cycle.  

Consider first a theory of massive particles.  In the deep IR,
the particles appear infinitely heavy and decouple,  leaving
an empty theory which cannot support a low-energy limit cycle. 
Therefore, we expect that a massive theory can only support
a UV limit cycle.  A massless theory on the other hand can
be non-trivial in the IR and thus support an IR limit cycle or
a limit cycle on all scales.  We will explore both possibilities
subsequently.

\subsection{RG equations for the S-matrix}

\def\Ecm{E_{\rm cm}}

It is well-known that the RG leads to certain scaling relations
for the correlation functions (Callan-Symanzik equations). 
Since the S-matrix is obtained from momentum space correlation
functions,  it thus also obeys RG scaling relations.  

Let $S$ denote the 2-particle to 2-particle S-matrix.  For an
integrable quantum field theory in 2 space-time dimensions, 
$S$ only depends on the kinematic variable 
$\Ecm^2 = (P_1 + P_2 )^2 $ where $P_{1,2}$ are the energy-momentum
vectors of the incoming particles. (For an integrable theory,
the incoming and outgoing momenta are the same.)  
Standard RG arguments applied to the S-matrix lead to the
scaling relation:
\beq
\label{II.2}
S( e^{-l} \Ecm , g(l_0 ) ) = S(\Ecm , g(l + l_0 )
\eeq 
(See e.g. ref. \cite{Ramond}.) 
For a theory with a limit cycle, when $l = \lambda$ the above
equation implies a periodicity in energy:
\beq
\label{II.3}
S( e^{-\lambda} \Ecm , g) = S(\Ecm , g ) 
\eeq

Let us now specialize to 2d kinematics.  First consider the
massive case,  where the energy momentum can be parameterized
in terms of a rapidity $\rap$:
\beq
\label{II.4}
E = m \cosh \rap , ~~~~~ p = m \sinh \rap 
\eeq
where $m$ is the mass of the particles.   The center of mass
energy is 
\beq
\label{II.5}
\Ecm^2 = 2 m^2 ( 1 + \cosh \beta), ~~~~~~~\beta = \beta_1 - \beta_2 
~~~~~({\rm massive ~ case}) 
\eeq
Suppose this theory has a UV limit cycle.   At high energies
$\beta$ is large and $\Ecm \approx m e^{\beta /2}$.   The relation
eq. (\ref{II.3}) then implies a periodicity in rapidity:
\beq
\label{II.6}
S(\beta - 2 \lambda ) = S(\beta )
\eeq

We turn now to the massless case.  Here the massless dispersion
relations $E = \pm p$ can be parameterized as:
\barray
E &=& \frac{m}{2} e^{\beta_R} , ~~~~~~ p = \frac{m}{2} e^{\beta_R}, 
~~~~~~~~~~~
{\rm for ~ right-movers} 
\nonumber
\\
E &=& \frac{m}{2} e^{-\beta_L} , ~~~~~ p = - \frac{m}{2} e^{-\beta_L}, ~~~~~~~
{\rm for ~ left-movers} 
\label{II.7}
\earray
where now $m$ is an energy scale.   The center of mass energy for
a right-mover with rapidity $\beta_R$ scattering with a
left-mover of rapidity $\beta_L$ is 
\beq
\label{II.8}
\Ecm = m e^{\beta /2} , ~~~~~~\beta = \beta_R - \beta_L ,
~~~~~~({\rm massless ~ case})
\eeq
If $S_{RL} (\beta)$ is the S-matrix for the scattering of 
right-movers with left-movers, then eq. (\ref{II.3}) again implies
a periodicity:
\beq
\label{II.9}
S_{RL} (\beta -  2 \lambda ) = S_{RL} (\beta ) 
\eeq

One sees then that the main difference between the massive and
massive case is that in the massive case the periodicity in
rapidity eq. (\ref{II.6}) implies a periodicity in energy $\Ecm$ 
only for large $\Ecm$, whereas in the massless case it leads to
a periodicity at {\it all} energy scales.  Thus a massless theory
with the periodicity eq. (\ref{II.9}) is consistent with 
a cyclic RG flow on all scales,  whereas the massive case 
at best can correspond to a theory with a limit cycle only in
the deep UV.

\subsection{RG equations for finite size effects}

Let us consider the quantum field theory 
 on a cylinder of finite circumference $R$
and length $L$ going to infinity.  
Viewing the finite circumference as the space the hamiltonian lives
on, and the length $L$ as the time,  the partition function
behaves as 
\beq
\label{4.1}
Z \sim e^{-L E(R)}
\eeq
where $E(R)$ is the ground state energy of the hamiltonian at
finite size $R$.  This ground state scaling function contains
a lot of information about the theory.  From $E(R)$ one conventionally
defines the effective central charge $c_{\rm eff}$:
\beq
E(R) = - \frac{\pi}{6} \frac{c_{\rm eff}(R)}{R}
\label{gs1}
\eeq

\def\ceff{c_{\rm eff}}

In theories with fixed points,  $c_{\rm eff} (R)$ is equal to the
Virasoro central charge $c$ at the fixed point, and smoothly 
interpolates between the ultra-violet  and infra-red 
values of $c$.   For this reason,  it is very interesting to
study $c_{\rm eff}$ in a theory  with no fixed points
but rather limit cycles.  

In order to derive an RG scaling equation for $\ceff$ we first relate
it to a correlation function.  
Let $\langle T_\mu^\mu \rangle $ denote the one-point function of
the trace of the stress-energy tensor on the cylinder.  It is a function of
the couplings $g$ and $R$, and is related to $\ceff$ as follows \cite{ZamoTBA}:
\beq
\label{cRG.1}
\langle T_\mu^\mu \rangle = \frac{2\pi}{R} \frac{d}{dR} \( R E(R) \) 
= - \frac{\pi^2}{3 R^2} \frac{d \ceff }{d\log R} 
\eeq
RG arguments applied to the one-point function give the scaling relation:
\beq
\label{cRG.2}
\langle T^\mu_\mu \rangle ( e^l R, g(l_0 ) ) 
= e^{-2 l } \langle T_\mu^\mu \rangle (R , g(l_0 + l ) )
\eeq
Taking $l$ to be equal to the period $\lambda$, one obtains:
\beq
\label{cRG.3}
\ceff' (e^\lambda R ) = \ceff' (R), ~~~~~
\ceff' \equiv \frac{d \ceff (R)}{d\log R} 
\eeq
i.e. $\ceff' (R)$ is a periodic function of $\log R$ with period $\lambda$.

\section{The models  and the cyclic RG flows}

In this section we describe an action for our  model 
and review the beta-function and RG flows.   Consider a conformal
field theory with $su(2)$ symmetry, described formally by
the action $S_{\rm cft}$.  The conformal symmetry promotes the
$su(2)$ symmetry to decoupled left-right current algebra 
symmetry \cite{KZ},  with left-moving (right-moving) currents $J^a (z)$  ($\bar{J}^a (\zbar)$),
$a=+,-,3$ where  $z=x+it, \zbar = x-it $ are  euclidean light-cone space-time
variables.    The currents are normalized to satisfy the following 
operator product expansion:
\beq
\label{2.1}
J^3 (z) J^3 (0) \sim \frac{k}{2 z^2} , ~~~~~
J^3 (z) J^\pm (0) \sim \pm \inv{z} J^\pm (0) ~~~~~
J^+ (z) J^- (0) \sim \frac{k}{2 z^2} + \inv{z} J^3 (0) 
\eeq
where $k$ is the level, and similarly for the right-moving currents.
  The simplest realization of the current
algebra is in terms of a pair of Dirac fermions, $J^a = \psi^\dagger
\sigma^a \psi $, where $\sigma^a$ are Pauli matrices, 
and corresponds to a $k=1$ reducible representation with
Virasoro central charge $c=2$.  Our model is based on the 
irreducible level $k=1$
current algebra which can be represented in terms of the 
left-moving part of a free boson 
$\Phi = \phi (z) + \bar{\phi} (\zbar)$:
\beq
\label{2.2}
J^3 =  \frac{i}{\sqrt{2}}  \d_z \phi, ~~~~~ 
J^\pm = \inv{\sqrt{2}} \exp \( {\pm i \sqrt{2} \phi}\) 
\eeq
The action $S_{\rm cft}$ is then just the action for a free massless
boson:
\beq
\label{2.3}
S_{\rm cft} = \inv{4\pi} \int  d^2 x  ~ \inv{2} (\d_\mu \Phi)^2 
\eeq
and leads to $\langle \Phi(x) \Phi(0) \rangle = - \log |x|^2 $. 
The above conformal field theory corresponds to Virasoro 
central charge $c=1$.  (In the condensed matter context, 
the $c=1$ theory can be obtained from the $c=2$ theory by
spin-charge separation.)

Our model is defined as an anisotropic left-right current-current
perturbation of the conformal field theory:
 \beq
\label{2.4}
S = S_{\rm cft} + \int \dx  \( 4 \gperp (J^+ \Jb^- + J^- \Jb^+ ) - 4 \gpar J^3
\Jb^3 \),
\eeq
where $\gperp, \gpar$ are marginal couplings. 
The RG beta-functions determine how the couplings $\gpar, \gperp$
depend on the length scale $e^l $.  
  These flows possess an RG
invariant $Q$ satisfying $dQ/dl  = 0$.  At one-loop, 
$Q=\gpar^2 - \gperp^2$.  In a particular renormalization
scheme, an all-orders beta-function has been conjectured \cite{GLM},
and the higher order corrections to $Q$ are  \cite{BLflow}:
\beq
\label{2.5}
Q = \frac{  \gpar^2- \gperp^2}{(1+\gpar)^2 (1-\gperp^2) }.
\eeq

The cyclic regime corresponds to $Q<0$.  Let us parameterize:
\beq
\label{2.6}
Q \equiv - \frac{h^2}{16} 
\eeq
with $h>0$.    The coupling $h$ is the main parameter of the
theory.  As shown in ref.  \cite{LRS},  it governs the period of
the RG cycles:
\beq
\label{2.7}
g( \lambda + l  ) = g(l),  ~~~~~ \lambda \equiv  \frac{2\pi}{h}
\eeq
where $g=g_{\parallel, \perp}$.  In each cycle 
$\gpar$ flows to $-\infty$ and jumps to $+\infty$.
As described in  \cite{BLflow} one can view the coupling
constant $- \arctan ( 4\gpar / h) $ as living on the universal cover of the 
circle.  The physical quantities, such as the S-matrix and 
the finite size effects are all finite and as we will see
reveal the RG cycles in a completely smooth manner. 
 The above periodicity is
not approached asymptotically,  but is present on all scales;
this will be an important criterion when we analyze the finite
size effects.

The limit $h \to 0$ corresponds to being on one of either of the two
separatrices $\gpar = \pm \gperp$, which can be taken in the 
half-plane $\gperp > 0$ due to the symmetry 
$\gperp \leftrightarrow - \gperp$.  These two $su(2)$ invariant
theories are quite different.  The theory with $\gpar = -\gperp < 0$
is a massive theory with UV fixed point at $\gpar = \gperp =0$,
and corresponds to the $su(2)$ invariant limit of the usual 
massive sine-Gordon theory with an empty IR fixed point.  
The theory with $\gpar = \gperp > 0$ on the other hand is a massless
theory with a non-trivial IR fixed point at $\gpar = \gperp = 0$.
In principle the UV behavior of the latter theory is not 
dictated by  the action eq. (\ref{2.4});  the action is rather
an effective low-energy theory.   However it is well understood
that this flow can arise from the $O(3)$ sigma model   
at $\theta = \pi$, which is defined on all scales, in particular
the UV.    A massless S-matrix for this flow was 
given in ref.  \cite{ZZflow}.  These observations lead us
to consider two different S-matrix theories for the region $h\neq 0$.
The first is the massive theory proposed in ref.  \cite{LRS} 
which when $h=0$ is the same as the sine-Gordon theory on the line
$\gpar = - \gperp < 0$.    The other theory,  not considered in  \cite{LRS}, 
is a massless theory which corresponds to the $O(3)$ sigma model 
with $\theta = \pi$ in the limit $h=0$. 
As argued in section II,  the massless theory 
is perhaps  more consistent with the RG flow since it has the signature
of a RG limit cycle on all length scales.

\subsection{Higher k}

In ref.  \cite{GLM} the beta-functions are given for arbitrary level $k$.  
The $k$-dependence is quite simple and can be scaled out of
the RG flows by defining a new couplings $\tilde{g} = k g$
and a rescaled RG time $\tilde{l} = l/k$.  In terms of $\tilde{g}$ 
and $\tilde{l}$ the flow equations are the same as for $k=1$.  Thus
the period in $l $ is  $2\pi k  /\tilde{h}$,
where $\tilde{h}$ is defined by $\tilde{Q} = - \tilde{h}^2 /16$,
with $\tilde{Q}$ the same as in eq. (\ref{2.5}) with $g$ replaced
by $\tilde{g}$.  However since the one-loop beta function does not
depend on $k$, it is more sensible to normalize   the RG invariant as :
\beq
\label{2.8}
Q = \frac{  \gpar^2- \gperp^2}{(1+k \gpar)^2 (1-k^2 \gperp^2) } 
  \equiv - \frac{h^2}{16}
\eeq
since then $Q = \gpar^2 - \gperp^2 + O(g^3)$ and 
the leading one-loop contribution to $Q$ is  the same for all $k$. 
 One sees that
$\tilde{h} = k h$,  and thus the period in $l$  is the same as for $k=1$:
$\lambda = 2\pi / h$ when $h$ is defined in the manner eq. (\ref{2.8}).

In the next section we will propose an  S-matrix for the higher $k$
theories.

\section{The S matrices }

\subsection{Massive case}

In reference  \cite{LRS} it was conjectured that the 
 cyclic regime of the KT flows described in the last section
has a spectrum consisting of   
a massive soliton and anti-soliton with topological charge $\pm 1$.
A factorisable S-matrix  was proposed based on the 
 $\CU_q (\hat{sl(2)})$  quantum affine symmetry of the 
underlying Hamiltonian
where  quantum parameter $q$  is related to the period $2 \pi/h$
of the RG cycles by the equation\footnote{In  \cite{LRS}, two
different S-matrices consistent with the quantum affine symmetry
were studied, differing by overall scalar factors.  In the
present article concerns only one of these S-matrices.  
The other S-matrix is characterized by a Russian doll spectrum
of resonances rather than a periodicity in the rapidity.}
\beq
q = - e^{- \pi h/2}
\label{s1}
\eeq

This value of $q$ is rather unusual for a relativistic factorized S-matrix 
theory.
The other previously  known models with S-matrices related to a quantum
affine symmetry all correspond to $q$ a  pure phase, and are models with
 fixed points corresponding to free bosons or their quantum group reductions
to minimal models when $q$ is a root of unity. 
S-matrices with a real $q$ however do appear in the
study of the XXZ spin chain in the massive antiferromagnetic regime, where
the solitons are understood  as massive spinons.  The relation of our model
to the XXZ spin chain was discussed in  \cite{LRS}. There it was shown
that in the low energy limit of small momentum, and as $h$ tends to zero, 
our S-matrix agrees with that of the spinons of the XXZ spin chain. 
However, in contrast to the spin chain, our dispersion relation is
relativistic on all energy scales, as appropriate to the relativistic
field theory of the last section, and this is the main difference between
the models.  As we will see, the interesting properties of the finite-size
effects which are a consequence of the cyclic RG only appear in the
ultraviolet limit for this massive case, and are thus unlikely to be 
properties of the spin-chain  since the latter 
 has an explicit lattice cut-off which leads to a non-relativistic 
dispersion relation.

We now describe in detail the S-matrix proposed in ref.  \cite{LRS}. 
As usual, we  let $\rap$ parameterize the energy and momentum of
the relativistic particles as in eq. (\ref{II.4}), 
where  $m$ is the mass of the soliton
and anti-soliton. To describe their scattering amplitudes let us introduce
the creation operators  $A_a (\rap )$ for the solitons/antisolitons,
where $a=\pm$ denotes the topological charge.  The S-matrix
may be viewed as encoding the exchange relation of these operators:
\beq
\label{s3}
A_a (\rap_1 ) A_b (\rap_2) = S_{ab}^{cd} (\rap_1 -\rap_2 ) ~ 
A_d (\rap_2 ) A_c (\rap_1 ) .
\eeq
The S-matrix can be presented in the following form:
\beq
\label{s4}
S (\rap ) =
\( \matrix{ S_{++}^{++} &0&0&0 \cr
0& S_{+-}^{+-} & S_{+-}^{-+} & 0 \cr 
0& S_{-+}^{+-} & S_{-+}^{-+} & 0 \cr
0&0&0 &S_{--}^{--} \cr 
} \) 
=  \frac{\rho (\rap )}{2i} 
\( \matrix{ q\spec - q^{-1}\spec^{-1}  &0&0&0 \cr
0& \spec - \spec^{-1}  & q-q^{-1} & 0 \cr  
0&q-q^{-1} & \spec - \spec^{-1} &0 \cr
0&0&0& q\spec - q^{-1}\spec^{-1}\cr
} \),
\eeq
where $q$ is given in eq. (\ref{s1}), and 
\beq
\label{s5}
\spec = e^{-i\rap h  / 2}.
\eeq
($\rap = \rap_1 - \rap_2 $.) 
The overall scalar factor is:
\beq
\label{s9}
\rho (q, \spec ) 
= \frac{2i q}{1-q^2 \spec^2 } 
\prod_{n=0}^\infty 
\frac{ (1 - q^{4+4n} \spec^{-2} )(1-q^{2+4n} \spec^2 )}
{ (1-q^{4+4n} \spec^2 )(1-q^{2+4n} \spec^{-2} )}
\eeq
This  infinite product is convergent since $|q| < 1$. 

The above S-matrix satisfies the necessary constraints of
crossing-symmetry, unitarity, real-analyticity and the
Yang-Baxter equation.  The factor $\rho$ can be understood
as the minimal solution to crossing and unitarity when 
$q$ is real.

The arguments leading the above S-matrix are entirely independent
of the RG analysis.  
  Indeed the S-matrix depends on  
the couplings $\gperp, \gpar$  only through the RG invariant $h$. 
The physical mass $m$ also depends on the couplings, though in
an unknown, non-perturbative fashion. 
It is therefore a non-trivial consistency check and confirmation  of the 
cyclic RG if the S-matrix possesses the cyclicities discussed
in section II.  Indeed, it does possess the cyclicity eq. (\ref{II.6}) 
with exactly the period $2\lambda$ predicted by the RG analysis
based on the all-orders beta-function.  This is easily
verified since $\spec ( \beta + 2 \lambda) = \spec (\beta)$.  
Recall that the period of the RG at one-loop is $h/\pi$ which is
$\lambda/2$ \cite{LRS}.  Thus, that the all-orders beta-function
correctly predicts the RG period is an indirect check of it.

S-matrices with real periodicities in the rapidity 
have previously been studied
in  \cite{Zelip,Mussardo}. Indeed, it was realized in 
 \cite{Zelip} that such a periodicity may imply an RG limit
cycle;  however at the time it was unknown  what field theory
the S-matrix described so it wasn't possible to do an RG analysis.
  In these  works, the S-matrix is expressed
in terms of elliptic functions,  whereas our S-matrix is 
trigonometric.   For a single particle with diagonal
scattering, crossing symmetry and unitarity  lead to  an imaginary 
$2\pi i$ periodicity,
so that an additional real periodicity in $\beta$ naturally leads to the
doubly periodic elliptic functions.  Our S-matrix on the other hand
is non-diagonal,  and thus is not $2\pi i $ periodic,  which explains
why it can be expressed in terms of trigonometric  functions.

One problem with the above  massive S-matrix has already been alluded to.
As discussed in section II, the above periodicity in rapidity 
implies an RG limit cycle only in the deep UV,  which contradicts
the flows described in section III which have limit cycles on
all length scales.   As we will see, other problems are encountered
when we investigate $c_{\rm eff}$.  We will argue that
  the massless S-matrix theory on the other
hand doesn't suffer from these problems.

For later purposes it is  convenient to express the component
$S^{+ +}_{+ + }$ as follows,
\beq
-i \log S^{+ +}_{+ + }(\rap) = \pi + 
\frac{\rap h}{2} + \sum_{n=1}^\infty \frac{2}{n} \frac{ \sin( n \rap h)}{
1 + e^{n \pi h}}
\label{s11}
\eeq
In the limit where $q \rightarrow -1$, i.e. $h \rightarrow 0$, it is easy 
to show that (\ref{s11}) becomes the $S^{+ +}_{+ + }$ entry
of the sine-Gordon model at the $SU(2)$ invariant point 
($SU(2)$ Thirring model):
\beq
-i \log S^{+ +}_{+ + }(\rap) = \pi + 
\int_o^\infty \frac{dx}{x} \sin(\rap x) \frac{e^{- \pi x/2}}{\cosh(\pi x/2)}  
\label{s12}
\eeq

\subsection{Massless case}

An S-matrix description of theories with IR fixed points was
given by Zamolodchikov and Zamolodchikov \cite{Zmassless,ZZflow}.  
An essential ingredient of their formulation are formal 
S-matrices for only left-movers or only right-movers, 
denoted $S_{LL}$ and $S_{RR}$.  These S-matrices are formal
S-matrices for a scale-invariant theory,  and encode the 
information, such as the Virasoro central charge,  of  the infra-red
fixed point.   When left-right scattering $S_{RL}$ is non-trivial,
the UV properties are affected.  For  the known models this collection
of  S-matrices corresponds to theories with non-trivial fixed points
in both the UV and IR.  

We now consider a massless  S-matrix description of the  limit cycle theory
described in section III.  Let us parameterize the massless
energy momentum for left and right movers as in eq. 
(\ref{II.7}).     Requiring the two-particle  S-matrices to correspond
to the $O(3)$ sigma model at $\theta = \pi$ in the $su(2)$ invariant
limit $h\to 0$ leads to the obvious proposal:
\barray
S_{RR} (\beta) &=& S_h (\beta ) , ~~~~~ \beta = \beta_{R1} - \beta_{R2}
\nonumber 
\\
S_{LL} (\beta) &=& S_h (\beta ) , ~~~~~ \beta = \beta_{L1} - \beta_{L2}
\label{masslessS}
\\
S_{RL} (\beta) &=& S_h (\beta ) , ~~~~~ \beta = \beta_{R1} - \beta_{L2}
\nonumber
\earray
where all the S-matrices $S_h$ on the right hand side 
are the same as in eq. (\ref{s4}).  

The above S-matrix again has the periodicity anticipated in
eq. (\ref{II.9}).   In this massless situation,  this is a
signature of cyclic RG on all length scales, entirely 
 consistent with the RG
flows.

\subsection{Higher $k$}

In this subsection we propose S-matrices for the higher level $k$ models. 
In the limit where $h = 0$,  the model is the well-known isotropic
current-current perturbation.  In the massive case 
($\gpar = - \gperp <0$) the particles are known to have 
RSOS kink quantum numbers in addition to the spin $1/2$  $su(2)$ quantum
numbers.  The S-matrix factorizes 
$S^{(k)} = S_{su(2)} \otimes S_{RSOS}^{(k)} $  where 
$S_{su(2)}$ is the $su(2)$ invariant S-matrix of the 
$su(2)$ Thirring model and $S_{RSOS}^{(k)}$ is the k-th RSOS
S-matrix which is a quantum group restriction of the usual
sine-Gordon S-matrix (restricted sine-Gordon model) and occurs
in the perturbations of the k-th  $c<1$ minimal conformal series \cite{ABL}. 
This factorization is a consequence of two commuting symmetries,
the $su(2)$ and the (fractional) supersymmetry \cite{ABL}.  

Based on the above $h=0$ result, it is straightforward to propose
an S-matrix for the k-th massive cyclic theory.   When $h\neq 0$, 
the $su(2)$ symmetry becomes the $\CU_q \( \hat{sl(2)} \)$ symmetry
with the same as above,   the fractional supersymmetry
is unaffected, and both symmetries continue to
commute. 
 The S-matrix is then:
\beq
\label{higherkS}
S^{(k)} =  S_h  \otimes S^{(k)}_{RSOS} 
\eeq
where $S_h$ is the S-matrix described in eq. (\ref{s4}),
and $S^{(k)}_{RSOS}$ is independent of $h$.

\section{Finite-size effects: massive case}

\subsection{Ground state scaling function}

As discussed  in section II, let 
us place  our model on a cylinder of finite circumference $R$
and length $L$ going to infinity, and consider the ground state
energy $E(R)$.   In this section we consider only the massive case.

Given the S-matrix one can derive integral equations for 
$\ceff$ by means of the thermodynamic Bethe ansatz (TBA) \cite{ZamoTBA}.  
The derivation of the TBA equations for diagonal S-matrices given
in  \cite{ZamoTBA}  makes no
reference to the usual Bethe ansatz for the eigenstates of 
a lattice regularization of the field theory, and relies only on knowledge of the S-matrix. 
   Unfortunately for non-diagonal S-matrices
the TBA normally leads to an infinite number of coupled integral
equations for the pseudo-energies.   However a dramatic simplification
to a single integral equation has been found by Pearce and 
Kl\"umper \cite{KlumperP}, 
and Destri and de Vega \cite{DDV}. In particular, 
Destri and de Vega (DdV) derived equations for the usual sine-Gordon
theory based on a special scaling limit of the XXZ spin chain.  
Since as described above it is in principle possible to derive
the TBA equations directly from the S-matrix alone, we can take
the point of view that DdV indirectly have dealt with the mathematical
complications of the non-diagonal
nature of the scattering in the sine-Gordon theory.   Since our
S-matrix is essentially a sine-Gordon S-matrix with a real $q$ 
rather than a phase,  we can thus obtain the TBA equations of
our model  from that of the sine-Gordon model equations by simply
using our expression for the S-matrix.


\subsection{The DdV equations}

In the approach developed by DdV, the effective central charge has
the following description in the usual sine-Gordon theory in the
regime with only solitons in the spectrum:
\beq
c_{\rm eff} (R) = \frac{6 m R}{\pi^2} \; {\imag} \; 
\int_{- \infty}^\infty
d \rap \; \sinh (\rap + i \eta)  \; \log( 1 + e^{ i Z(\rap + i \eta)})
\label{gs4}
\eeq
where $m$ is the soliton mass and  
$Z$ is a solution of the non-linear integral equation:
\beq
Z(\rap) = m R \sinh \rap + 2\;  {\imag} \int_{- \infty}^\infty
d \rap' \; G(\rap - \rap' - i \eta) \; \log( 1 + e^{ i Z(\rap' + i \eta)})
\label{gs2}
\eeq
The kernel $G(\rap )$ depends only on the soliton to soliton
S-matrix element:
\beq
G(\rap) = \frac{1}{2 \pi i} \frac{\partial}{\partial \rap} \log 
S^{+ +}_{+ +}(\rap)
\label{gs3}
\eeq
Above,  $0<\eta<\pi/2 $, is  a regulator needed for convergence.  

For the usual sine-Gordon model with $S^{++}_{++} $ given in
ref.  \cite{ZZ},  $\ceff (mR)$  is an uneventful function that
smoothly interpolates between $c_{UV} = 1$ in the ultra-violet ($R \to 0$)
and $c_{IR} = 0$ in the infra-red ($R \to \infty$). 
This reflects the existence of a $c=1$ UV fixed point, a free
massless boson, and an empty IR fixed point.  The fact that
$c_{IR} = 0$ is simply due to the fact that the theory is massive
and in the deep IR these massive states decouple.  
This is in accordance with the rough picture that $\ceff$ is
in a sense a measure of the massless degrees of freedom:  in the
deep UV the massive particles are effectively massless and $c_{UV} = 1$,
whereas in the deep IR the massive particles decouple and 
$c_{IR} = 0$.    We have verified the above statements by solving
the integral equation (\ref{gs2})  numerically for the usual sine-Gordon
model.  The UV and IR limits of $\ceff$ can also be computed
analytically in terms of dilogarithms \cite{ZamoTBA,DDV}.

Because of the cyclic nature of the RG in our model, we expect 
a much more interesting behavior in the UV due to the non-existence
of a fixed point.   In the IR on the other hand, the usual argument
given above applies since the theory is massive and one concludes
$c_{IR} = 0$.   

 For our model,  the kernel has the simple
expression:
\beq
G(\rap) =
\frac{h}{4 \pi} \left[1 
 + 4 \sum_{n=1}^\infty  \frac{ \cos( n \rap h)}{
1 + e^{n \pi h}} \right].
\label{gs5}
\eeq
which follows from eqs. (\ref{gs3},\ref{s11}).  
The kernel has the following periodicity:
\beq
\label{Gperiod}
G\( \rap + \frac{2\pi}{h} \) = G(\rap )
\eeq
which is a clear indication of the cyclic RG, with the period
exactly as predicted by the RG analysis reviewed in section III.  
It is this cyclicity of the kernel that is ultimately responsible
for the non-trivial behavior of $\ceff$ in the UV.  

Let us first rewrite the DdV equations so that they resemble the
more
conventional expressions \cite{ZamoTBA}, and also make them
more suitable for numerical solution.   Since $Z(\rap )$ is
a real function when $\rap$ is real,  let us define the 
pseudo-energies 
\begin{equation}
\vep(\rap) = -iZ(\rap + i \eta), \qquad
\vep^*(\rap) = iZ(\rap - i \eta), 
\label{gs7}
\end{equation}
Then eq.  (\ref{gs2}) can be written compactly as 
 \beq
\vep = \vep_0 - G_0 * L + G_1 * L^*,
\label{gs13}
\eeq
where $*$ denotes a convolution
\beq
(f*g)(\beta ) = \int_{- \infty}^\infty \; d \beta' \;
f(\beta - \beta') \; g(\beta').
\label{gs14}
\eeq
and we have defined:
\barray
\vep_0(\rap)&  = & -i m R \sinh (\rap + i \eta),
\label{gs10} \\
L(\rap) & = & \log( 1 + e^{-\vep(\rap)}),
\label{gs11} \\
G_0(\rap) & = & G(\rap), \qquad
G_1(\rap) \ = \ G(\rap+ 2 i \eta),
\label{gs12}
\earray
The expression for $\ceff$ becomes
\beq
\ceff (mR) = \frac{6}{\pi^2} \; 
\real \int_{- \infty}^\infty
d \rap ~   \vep_0(\rap) L(\rap)  
\label{gs15}
\eeq

We have solved the above equations numerically and 
verified that $\ceff$ is independent of $\eta$.  
A detailed presentation of our numerical results will
be given below.  For the further development of analytic results,
the value $\eta = \pi/2$  is preferred since then $\vep_0 = 
mR \cosh \rap $ is the standard pseudoenergy for a free theory,
 and when the kernel $G$ is zero,
which occurs at the free fermion point of the usual sine-Gordon
model, one has $\vep = \vep_0$ and 
 the expression for $\ceff$ is the standard free
fermion one:  
\beq
\label{cfree}
c_{\rm free} (mR) = \frac{6 mR}{\pi^2} \int_{-\infty}^\infty 
d\rap \cosh \rap ~ \log ( 1 + e^{-mR \cosh \rap } ) 
\eeq
In fact,  in the usual sine-Gordon theory, $\ceff$ is
rather well approximated by $c_{\rm free}$ at all values of the coupling.  
Henceforth $\eta$ will be set to $\pi /2 $ in all analytic
computations, and unless otherwise stated:
\beq
\label{epzero}
\vep_0 (\rap ) = mR \cosh \rap .
\eeq

It is convenient to define the function
\beq
\label{4.12}
\sigma (\rap ) \equiv \vep (\rap) - \vep_0(\rap).
\eeq
Since the kernels $G_{0,1}$ are periodic functions of
$\rap$ with period $2\pi /h$, then so is 
$\sigma(\rap )$, and it can thus be expanded as a Fourier
series:
\beq
\label{4.13}
\sigma (\rap ) = \sum_{n=-\infty}^\infty  \sigma_n ~ e^{i n h \rap } 
\eeq
We can then proceed to rewrite the integral equations as non-linear
algebraic equations for the various Fourier coefficients.  First
we expand the kernels:
\barray
G_0 (\rap ) &=& \frac{h}{2\pi} \sum_{n=-\infty}^\infty g_{0,n} 
e^{inh\rap} , ~~~~~
g_{0,n} = \inv{1 + e^{|n| \pi h}}  \label{4.14}  \\
G_1 (\rap ) &=& \frac{h}{2\pi} \sum_{n=-\infty}^\infty g_{1,n} 
e^{inh\rap} , ~~~~~
g_{1,n} = e^{-n \pi h} g_{0,n}.
\nonumber
\earray
Then the integral equation (\ref{gs13}) can be written as
\beq
\label{4.16}
\sigma_n = - g_{0,n} L_n + g_{1,n} L_{-n},
\eeq
where we have defined
\beq
\label{4.17}
L_n = \frac{h}{2\pi} \int_{-\infty}^\infty d\rap 
~ e^{-inh\rap} L(\rap ),
\eeq
with $L( \rap )$ as before but now viewed as a function of
$\sigma$.   In deriving eq. (\ref{4.16}) 
we have used that $(L_n)^*  = L_{n}$,  which is
a consequence of $(L (\beta))^* = L(-\beta)$.
From equation (\ref{4.16}) one can derive the following
relation:
\beq
\label{4.18}
\sigma_n = - \frac{g_{1,n}}{g_{0,n}} \sigma_{-n} = - e^{-n \pi h} 
\sigma_{-n}.
\eeq

\no which implies the absence of the zero mode, i.e.
$\s_0 =0$.

\subsection{Exact Numerical results}

The numerical work is simplified  
by splitting the various 
functions in the model, i.e. $f= \vep, \sigma, L$, 
and the DdV equations into their real and 
imaginary parts (e.g. $f = f'+ i f''$). Moreover 
the symmetry properties satisfied by these quantities
further reduce the computational resources of the program. 
The kernel $G(\rap)$ in eq.~(\ref{gs5}) is an 
even function of $\rap$, and 
$Z(\rap)$ in eq.~(\ref{gs2}) is a real quantity. These facts imply that 
$Z(\rap)$ is an odd 
function of $\rap$, which in turn 
determines the parity properties of the remaining
functions in the theory.  
Namely, the 
real parts, $f'(\rap)$, are even, while the imaginary 
parts, $f''(\rap)$, are odd functions of $\rap$, and 
therefore $f(-\rap) = f^*(\rap)$.

The solution of the 
DdV equations has been found using two different 
methods: 1) recursive solution of the 
eq.~(\ref{gs13}) for the functions $\sigma'(\rap)$ 
and $\sigma''(\rap)$;  
and 2) finding the roots of the non-linear 
system ~(\ref{4.16}) for the modes $\s_n$
subject to the constraint (\ref{4.18}). 
The results obtained by  these two methods are in full agreement.

\begin{figure}[h!]
\begin{center}
\includegraphics[height=10.0 cm, angle=0]{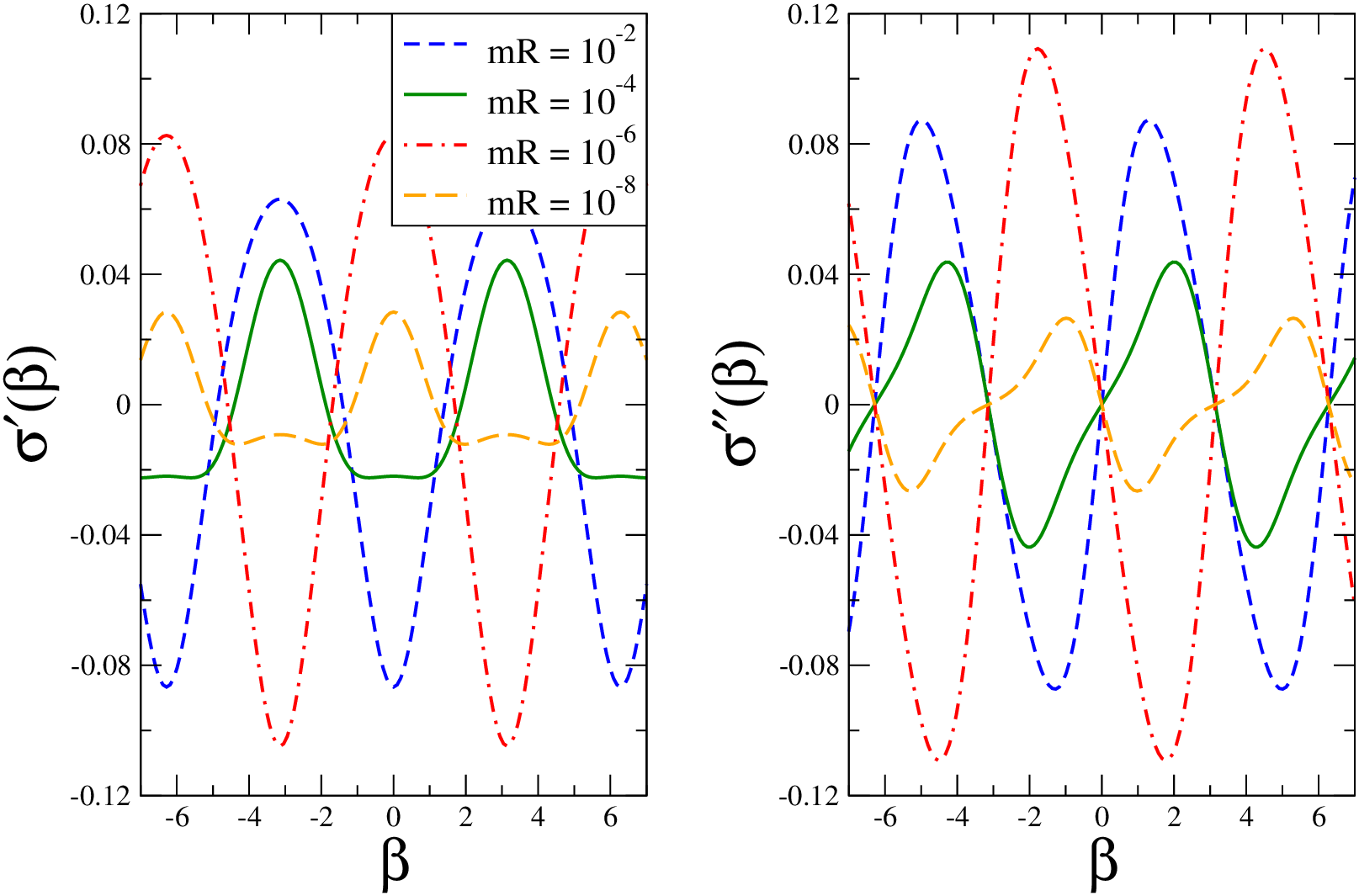}
\end{center}
\caption{Solution of the eqs.(\ref{gs13}) for $\s'(\rap)$
and $\s''(\rap)$ for $h=1$, where 
$\s= \vep - \vep_0=\s'+ i \s''$.} 
\label{sigma10}
\end{figure}

In fig \ref{sigma10} we show the solution to eq.~(\ref{gs13})
in the case where $h=1$ and several values of $m R$ ranging
from $m R = 10^{-2}$ to $10^{-8}$. These results confirm that 
$\s'$ is an even and periodic function of $\rap$ 
with period $2 \pi/h$, 
while $\s''$ is an odd periodic function with the same period. 
We also observe that the amplitude of the oscillations is rather
small, of order $10^{-2}$ in this case, which is a general feature 
except for special regions of $mR$ where the amplitudes are of order 1.

\begin{figure}[t!]
\begin{center}
\includegraphics[height=10.0 cm, angle=0]{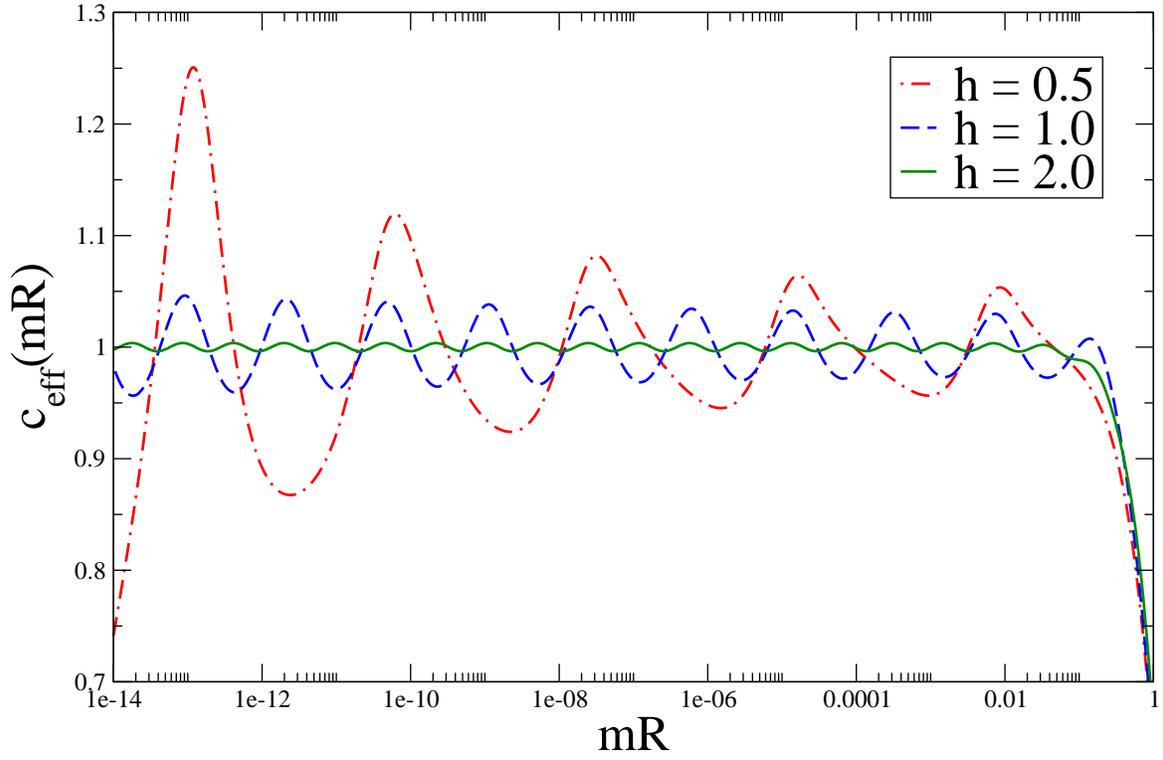}
\end{center}
\caption{Exact numerical values of $\ceff(mR)$ for $h=0.5,1,2$
in the range $mR = 1-10^{-14}$. 
} 
\label{c-oscillatory}
\end{figure}

Fig.~\ref{c-oscillatory} shows the values of $\ceff(m R)$ corresponding
to $h=0.5$, $1$, $2$ in the  range $mR = 1-10^{-14}$. 
In the IR region  $\ceff$
goes to $0$ exponentially fast and independently of the value of $h$, 
as expected for  a massive model. 
As soon as $mR \lesssim 10^{-2}$, $c_{\rm eff}$ starts  to oscillate around 
$c=1$ with a period $\pi/h$  in the scale, i.e.,
\beq
c(m R) \sim  c(e^{\pi/h} m R)
\label{num7}
\eeq
Indeed, as seen in the figure, a complete cycle 
for the  $h=0.5$ model contains two and four cycles 
of the $h=1$ and $h=2$  models respectively.
Another feature of $c_{\rm eff}$ 
is that the amplitudes
of the oscillations increase to-wards the UV, which is  especially 
pronounced for 
$h=0.5$,  and this is why eq.~(\ref{num7}) is not an 
equality. The RG analysis performed in section III  
suggests a periodic $\ceff(m R)$ function 
with period $2 \pi/h$ in the scale, and this  is twice 
what is observed numerically.  This will be explained analytically
in the next subsection.

\subsection{Approximate analytic solution}

In this subsection we derive an approximate analytic expression 
for $\ceff$, explaining the properties observed numerically. 
Later we will show that it agrees remarkably
well with the exact numerical solution to the full non-linear problem,
and arguments for why this is so will be given.  

\def\Kbar{K'}

We first make a linear approximation to the integral equation 
(\ref{4.16}).   Using 
\beq
\label{4.19}
L (\rap ) = L_0 (\rap ) - K (\rap ) \sigma (\rap ) + \CO (\sigma^2 ) 
\eeq
where 
\beq
\label{LK}
L_0 (\rap ) = \log (1 + e^{-\vep_0 (\rap )} ) , ~~~~~
K (\rap ) = \inv{1 + e^{\vep_0 (\rap )} }
\eeq
then to order $\sigma$:
\beq
\label{4.20}
L_n = L_{0,n} - \sum_m K_{n-m} \sigma_m 
\eeq
where $L_{0,n}$ and $K_n$ are Fourier transforms of $L_0$ and $K$:
\beq
\label{KLn}
L_{0,n} = \frac{h}{2\pi} \int_{- \infty}^\infty  
d\rap \; e^{-inh\rap} L_0 (\rap ) ,~~~~~ 
K_n  = \frac{h}{2\pi} \int_{- \infty}^\infty 
d\rap \; e^{-inh\rap} K  (\rap )
\eeq
Using eq. (\ref{4.20}) in eq. (\ref{4.16}) one finds the linear
equations:
\beq
\label{siglin}
\sum_m  A_{n,m} \sigma_m = B_n 
\eeq
where 
\begin{eqnarray}
A_{n,m} &=&\delta_{n,m} - g_{0,n} K_{n-m} + g_{1,n} K_{-n-m}
\label{4.23} \\
B_n &=&  g_{1,n} L_{0,-n} - g_{0,n} L_{0,n} 
\nonumber 
\earray
In this approximation,
\beq
\label{4.24}
\ceff (mR) \approx  c_{\rm free} (mR) - \frac{12}{\pi h} 
\sum_n  \sigma_n \Kbar_n 
\eeq
where $\Kbar_n $ is the following Fourier transform:
\beq
\label{4.25}
\Kbar_n = 
\frac{h}{2\pi} \int_{- \infty}^\infty 
d\rap \; e^{-inh\rap} \Kbar  (\rap ), ~~~~~~
\Kbar  (\rap ) \equiv \vep_0 (\rap ) K (\rap ) 
\eeq

All of the needed Fourier transforms $L_{0,n}, K_n , \Kbar_n $ are
computed in appendix A.   There the $mR$ dependence is expressed
in terms of the logarithmic variable
\beq
\label{sdef}
s = \log \( 2\pi /mR \) 
\eeq
In the deep UV where  $R$ goes to $ 0$, $s$ is very large. 



From the results in appendix A one finds in the UV limit of large $s$:
\barray
K_0 &\approx& \frac{h}{2\pi} ( s - \gamma - \log 2)
\label{4.27} 
\\
K_{n\neq 0} &\approx&  -\frac{h}{\pi \cosh (n\pi h/2 ) } 
{\real } \( e^{inhs} \zeta_1 (1-i n h) \)
\nonumber
\earray
where $\gamma$ is Euler's constant and 
\beq
\label{zetadef}
\zeta_1 (z) \equiv (1- 2^{-z} ) \zeta (z) 
\eeq
with $\zeta$ Riemann's zeta-function.   Thus $K_0$ is large
compared to $K_{n\neq 0}$ and this justifies the further
approximation of keeping only the $K_0$ terms in eq. 
(\ref{siglin}).  ($K_{n\neq 0}$  is further suppressed by 
$e^{-n\pi h/2}$.)   Using eq. (\ref{4.18}), this essentially
diagonalizes eq. (\ref{siglin}) and one finds
\beq
\label{4.28}
\sigma_n \approx \frac{B_n}{1- 2 g_{0,n} K_0 } 
\eeq
Using eq. (\ref{4.28}) and the results in  appendix A in 
eq. (\ref{4.24}) one finds 
\beq
\label{clinear}
\ceff (mR) = c_{\rm free} (mR) + \frac{24}{\pi} 
\sum_{n=1}^\infty 
\frac{ {\imag} \( e^{2i n s h} \zeta_1^2 (-i n h) \) }{
n(1 + e^{n\pi h} - \frac{h}{\pi} (s-\gamma - \log 2 )) }
+ \CO( e^{-2s} ) 
\eeq
which is meant to be valid in the UV.  The contribution 
$c_{\rm free}$ rapidly approaches $1$ in the UV.  

A direct consequence of this equation, together with
(\ref{sdef}), is that the amplitudes of the
oscillations of $\ceff(mR)$ increases to-wards the UV
because of the presence of a pole at the
critical distance $R_{c,n}$  
\beq
mR_{c,n} = \pi e^{-\gamma } \; \exp \left( 
- \frac{\pi}{h} ( 1 + e^{n \pi h}) \right)
\label{per17}
\eeq
One sees that  $\ceff(mR)$ becomes singular at 
$R_{c,n}$  in this  linear approximation.
The double exponential in (\ref{per17}) implies
that these scales are extremely  small. For example
the highest value of $R_{c,1}(h)$ is attained at
$h= 0.407$ where $mR_{c,1}= 7.14 \; 10^{-16}$.
Therefore at this scale we expect our approximation
to break down. In the next subsection we shall
study in more detail these critical points.

In fig.~\ref{linear08} we compare the result (\ref{clinear}) with the exact numerical
one for $h=0.8$ in the range $mR =1 - 10^{-20}$, observing 
a very good agreement over  a wide range of distances. 
In this case the  first critical distance 
 is given by $mR_{c,1} = 3.06 \; 10^{-23}$.
Eq.~(\ref{clinear}) explains the oscillatory behavior of $\ceff (mR)$
with a period $\pi/h$ in the scaling variable $s$. The lack of 
periodicity is due to the denominator of this expression which
becomes more important as one  approaches the critical points $R_{c,n}$.
  
The reason why the linear approximation is so good is that
the higher order terms in the expansion of $L$
are in fact strongly suppressed in the deep UV 
(see Appendix B for details).
 
In fig.~\ref{linear04} we give a similar comparison for $h=0.4$
over a range  that contains two critical distances 
$mR_{c,1}= 7.09 \; 10^{-16}$ and $mR_{c,2} = 5.33 \; 10^{-46}$.

Finally consider the $h\to 0$ limit, where the period $\lambda$ 
becomes infinitely large.  The singularities at $R_{c,n}$ are
pushed deeper and deeper to the UV and approach zero.   One thus 
approaches a smooth flow between $c_{UV} =1$ and $c_{IR} = 0$ 
as expected for the isotropic limit of the massive sine-Gordon theory.

A curious property of eq.(\ref{clinear}) 
is that for the special value $h= h_1 = 2 \pi/\log 2$
the linear contribution to $\ceff$ vanishes 
as a consequence of the identity $\zeta_1(i n h_1) =0, \;
\forall n$. Thus the value of $\ceff$ is given simply
by $c_{\rm free}$. The exact numerical solution also
satisfies this result to order $10^{-16}$ which was our  
machine precision. For $h$ nearby $h_1$ 
the value of $\ceff - c_{\rm free}$ 
is small but never zero. 
This fact is a  consequence
of the ``absence'' of the prime number 2 in  Euler's
product formula for  the zeta function
$\zeta_1(z)$. For values of $h \neq h_1$ the corrections to $c_{\rm free}$
in eq. (\ref{clinear}) are  never zero since
$\zeta(ih)$ does not vanish.  
The later quantity is related by the 
Riemann's functional equation to 
$\zeta(1 - i h)$ which does not vanish as well. The latter
result was used by 
Hadamard and de la Vall\'ee-Poussin to prove the prime number theorem
 \cite{zeta}.

In summary,  in the massive case the effective central charge 
$\ceff$ exhibits nearly periodic structures over a large
range of scales in the ultra-violet, with a period consistent
with the RG predictions.   However as one
approaches the  scales $R_{c,n}$ in the very deep ultraviolet, 
$\ceff$  develops a singularity and the periodicity eq. (\ref{cRG.3}) 
is violated.  In the analytic approximation this singularity
appears as a pole in $s$ in the equation (\ref{clinear}).  
These features suggest that the massive theory is not
entirely consistent with the RG flows, as alluded to in section IV.

\begin{figure}[t!]
\begin{center}
\includegraphics[height=10 cm, angle=0]{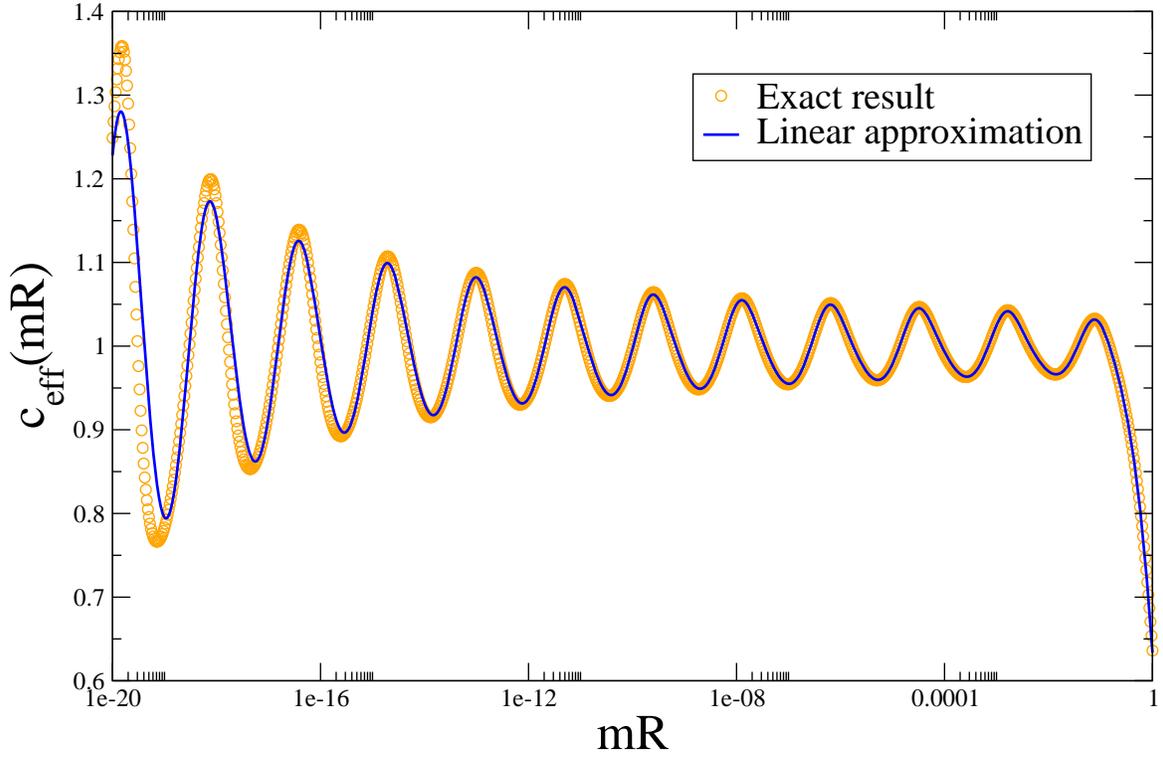}
\end{center}
\caption{Comparison of $\ceff(mR)$ computed with the 
analytic approximation eq.(\ref{clinear}) and the 
exact numerical results for $h=0.8$}
\label{linear08}
\end{figure}

\begin{figure}[t!]
\begin{center}
\includegraphics[height=10 cm, angle=0]{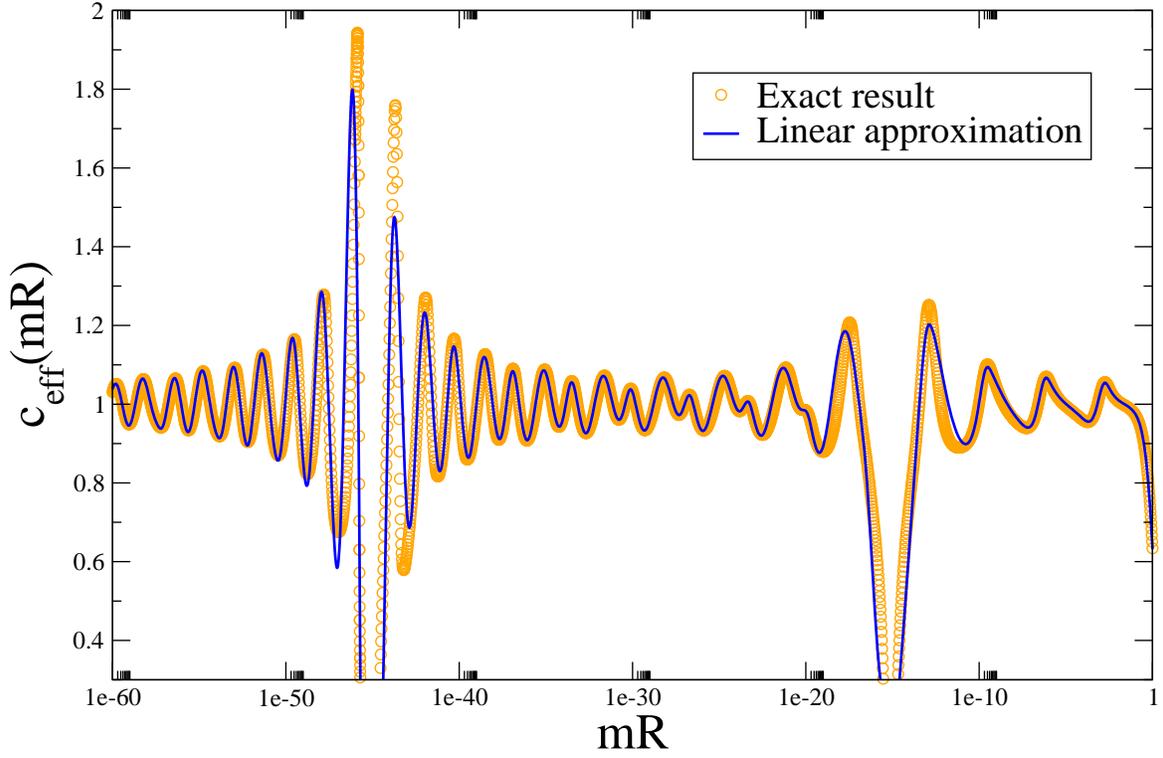}
\end{center}
\caption{Comparison of $\ceff(mR)$ computed with the 
analytic approximation eq.(\ref{clinear}) and the 
exact numerical results for $h=0.4$}
\label{linear04}
\end{figure}

\subsection{Coexistence regions}

The approximate analytic solution of the DdV equations presented
above explains the main features of the behavior
of $\ceff$, namely its periodicity and scale dependent
amplitude. For most values of $mR$ the mode amplitudes
$\s_n$ are rather small for all $n$, which explains the high accuracy
of the analytic approximation. However at the critical values
$R_{c,n}$ the $n^{\rm th}$ mode blows up and  
the analytic approximation is no longer valid, meanwhile the 
numerical solution still gives finite results. 

\begin{figure}[t!]
\begin{center}
\includegraphics[width=11.5 cm, angle=0]{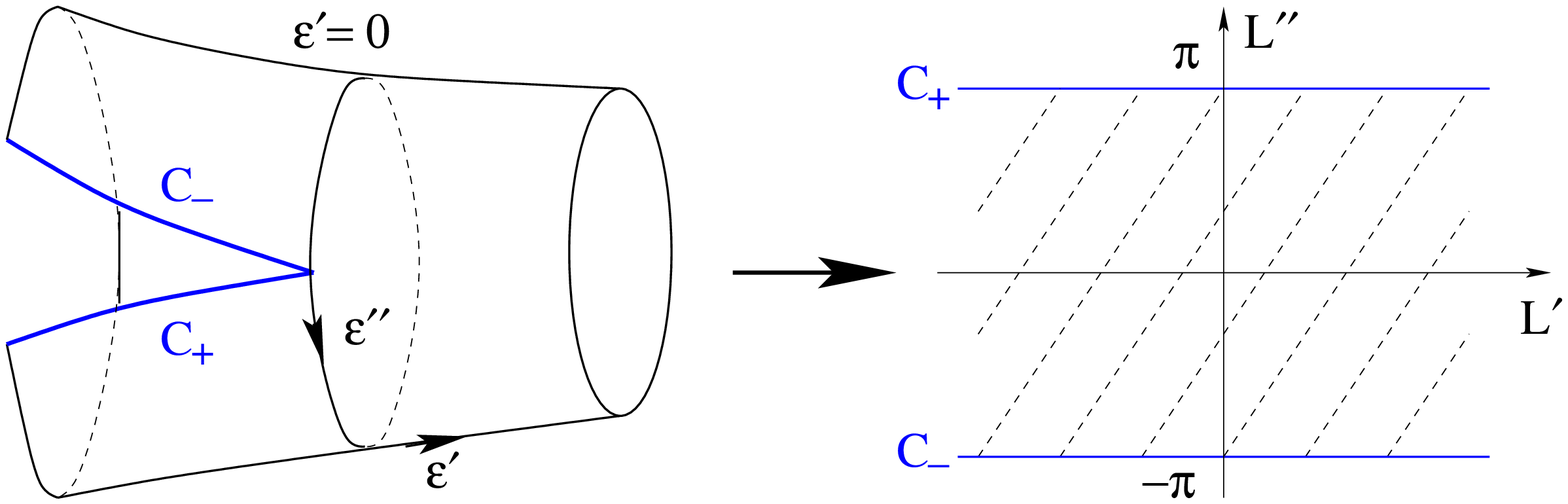}
\end{center}
\caption{Conformal map $L(\vep) = \log(1 + e^{-\vep})$ from
the cylinder with a cut at  $\vep' < 0$ and $\vep'' = (2 p +1) \pi$
into the strip $- \pi < {\imag} L(\vep) < \pi$. 
} 
\label{cut}
\end{figure}
%
Before we focus on this problem let us consider
an important technical point regarding  
the definition of the logarithm $L(\rap)$ in (\ref{gs11}).  
The imaginary part of $L(\rap)$  is the phase 
of $w(\rap) = 1+e^{-\varepsilon(\rap)}$.  
For $-\infty < \rap < \infty$ the variable  
$w(\rap)$ describes 
a trajectory in the complex plane, which may pass  
through the cut of the logarithm, changing branches.
 As long as the modes $\s_n$ remain small 
enough this never happens. 
We have found that the solutions of the DdV eqs. belong to  
the principal branch of the logarithm, i.e.  
 $-\pi < {\imag }L(\rap) < \pi$. Under this condition 
$L(\vep)$ becomes a conformal map from the 
cylinder with a cut at $\vep' < 0$ and 
 $\vep'' = (2p+1)\pi$,
into the strip  $-\pi < {\imag }L(\vep) < \pi$, 
as shown in fig.~\ref{cut}. The cut corresponding to $p=0$ 
translates into, 
\begin{eqnarray}
\vep' & = & \vep'_0 + \sum_{n=1}^{\infty}\s'_n \cos(nh\rap) \ < \ 0, 
\qquad
\s'_n = -\tanh(nh\alpha)\s''_n,
\nonumber \\
\vep'' & = & \vep''_0 + \sum_{n=1}^{\infty}\s''_n \sin(nh\rap) \ = \ \pi,
\end{eqnarray}
and for $\alpha = \pi/2$ ($\vep''_0 = 0$) 
these equations simplify.  Since just 
one mode dominates the behavior of the system near the corresponding
singularity, 
let us  consider the single mode equations, 
\begin{eqnarray}
\vep' & = & mR \cosh\rap + \s'_n \cos(nh\rap) \ < \ 0, 
\qquad
\s'_n = -\tanh(nh\pi/2)\s''_n,
\nonumber \\
\vep'' & = & \s''_n \sin(nh\rap) \ = \ \pi,
\end{eqnarray}
which implies that $|\s''_n| < \pi$. 
Now we can solve the eqs.~(\ref{4.16}) for a 
single mode and search for the values of $R$ around 
$R_{c,n}$ satisfying $\s''_n = \pm\pi$.  
The  results are shown in fig.~\ref{coexistence}.

This study of the logarithm solves the problem 
of the unbounded growth of $\s''_n$ and 
consequently $\ceff(mR)$. In the IR there is a unique
solution which evolves
continuously to-wards the UV 
until it hits the cut, i.e. $|\s''_n| = \pi$,
beyond which there is no solution.
The boundary of the ``IR phase'' has a sawtooth shape
which on average follows the analytic estimate 
(\ref{per17}) (see Fig.~\ref{coexistence}). 
For smaller values of $R < R_{c,1}$ one finds again
a unique UV-solution which can be smoothly continued
to-wards the IR until it hits the cut again. The boundary
of this UV-solution has also a characteristic sawtooth shape
as shown in Fig.~\ref{coexistence}. The IR and UV phases
may coexist, meaning that there are two solutions for the same
values of $mR$ and $h$.
The regions with no IR/UV coexistence may 
contain isolated solutions non connected to the 
IR or UV regions, or they may not  contain solutions at all.
We have encountered  both kinds of situations.The previous discussion concerns the truncation of the
DdV equation to a single mode. Including all the modes
yields an  infinite number of coexistence regions, one
for every mode, as shown already by the analytic approximation. 
The whole picture is as follows. Moving from the IR to-wards
the UV, the first mode $\s''_1$ passes through its critical 
region and gets damped beyond it. Then, the second mode
takes over and after a crossover region the period
of $\ceff$ doubles. This second mode also reaches
its critical region and the process repeats itself 
with the third mode ( see fig.~\ref{linear04}). 
Once more the analytic result fits amazingly well
the exact numerical result including the proximity 
of the coexistence regions (for clarity 
we have truncated the analytic part very close to  
the pole).

\begin{figure}[t!]
\begin{center}
\includegraphics[height=9.9 cm, angle=0]{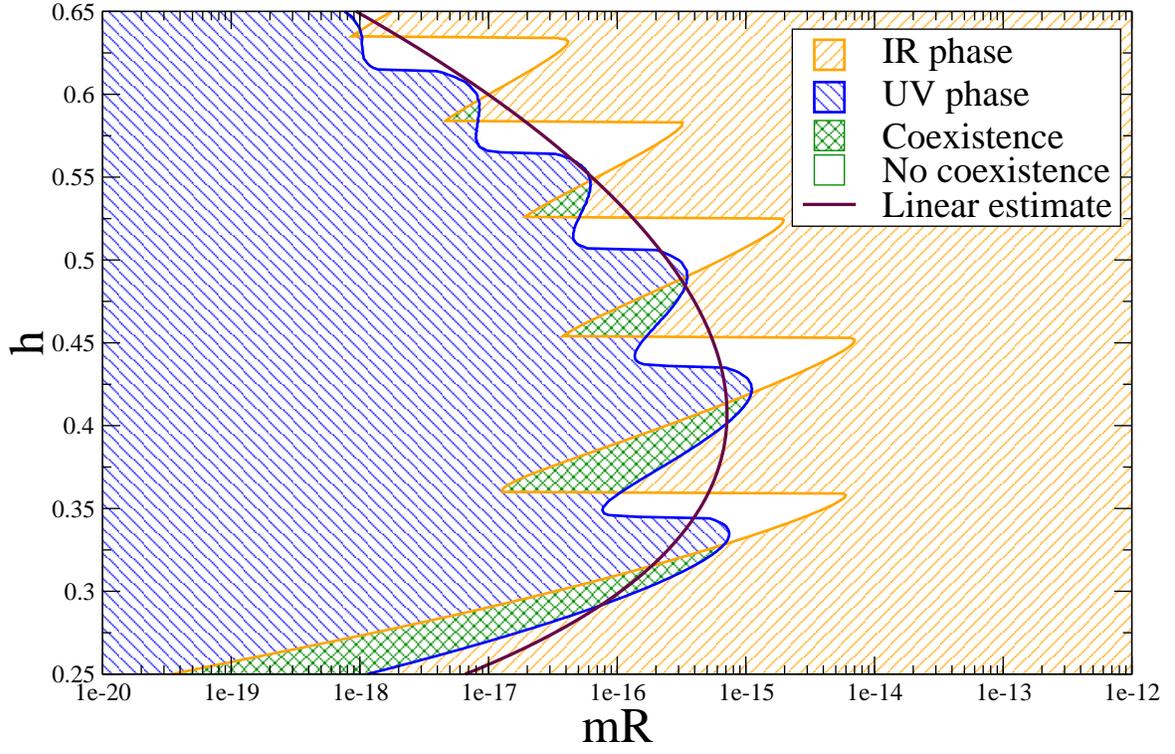}
\end{center}
\caption{Coexistence of the IR and UV phases.}
\label{coexistence}
\end{figure}
\begin{figure}[h!]
\begin{center}
\includegraphics[height=9.9 cm, angle=0]{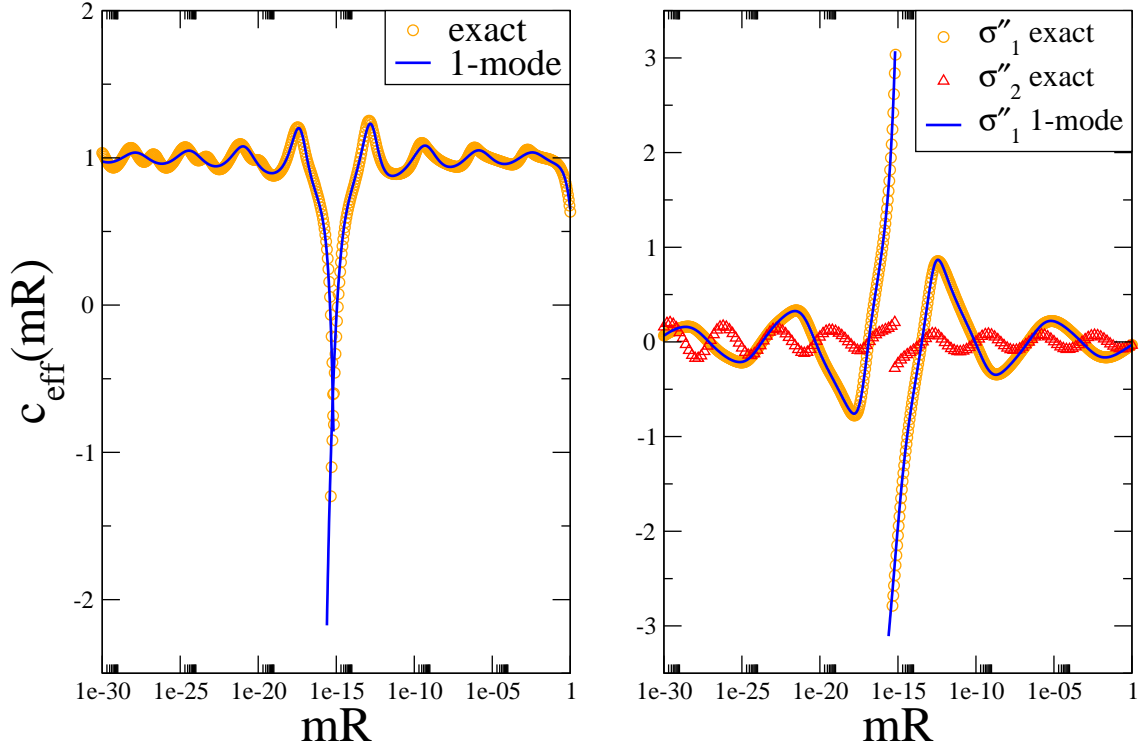}
\end{center}
\caption{$\ceff(mR)$ and the $\sigma'_1$ and $\sigma'_2$ 
in the region of the ''singularity'' for $h=0.4$.}
\label{sgmod04}
\end{figure}

\vspace{1cm} Finally, fig.~\ref{sgmod04} presents a comparison of 
the exact solution and the single mode approximation 
near the first singular region. 
At  $h = 0.4$ there is IR/UV coexistence 
as can be checked in fig.~\ref{coexistence}.  
Hence the IR and UV curves $\ceff(mR)$ 
overlap around $mR \sim 10^{-15}$ (fig.~\ref{sgmod04}a). 
Fig.~\ref{sgmod04}b shows that the first mode of the exact
solution, $\s''_1$, dominates over the second mode,    
$\s''_2$, which in the UV becomes more relevant.
Notice also that $\s''_1$ is very well described
by the single mode approximation. 

\section{Remarks on $\ceff$ for the   massless case}

The  structure of the TBA equations for massless theories
has been understood in a number of models \cite{Zmassless,fateev,ZZflow}.
Unfortunately the massless DdV form of the equations  are not known
for the $O(3)$ sigma model at $\theta = \pi$, thus unlike the
massive case,  it is unknown what the $h=0$ limit of the TBA equations
based on the S-matrices in eq. (\ref{masslessS}) should be.  
In spite of this, based on the known structure of massless TBA equations
we can anticipate the general properties $\ceff$ for theories based
on massless S-matrices with limit cycles.

When the LR scattering is trivial,  the S-matrices $S_{LL}$ and 
$S_{RR}$ describe a conformally invariant theory. 
Let us introduce pseudo-energies $\vep_L , \vep_R$ separately for the 
left and right movers.    As for other models, 
when the LR scattering is trivial the TBA equations are 
two decoupled integral equations for $\vep^{L}$ and $\vep^R$, 
each equation having 
the same form as a massive theory,  the only difference being in 
the source terms $\vep^L_0, \vep^R_0$ 
 which now reflect the massless dispersion
relations.  Thus,  when the left-right scattering is turned off,
the TBA equations should just be two decoupled DdV equations of the kind
described in section V. 
  Turning on the left-right scattering $S_{LR}$  then couples
the L,R integral equations.

Consider a toy model with only one particle with $S_{LL} = S_{RR} = 1$
but a non-trivial $S_{LR}$, and define as usual the kernel as
\beq
\label{masslessG}
G(\beta ) = \inv{2\pi i} \d_\beta \log S_{LR} 
\eeq
An example is the flow between the tricritical and critical Ising
model \cite{Zmassless}.  
Define the free pseudoenergies:
\beq
\label{masslessTBA.1}
\vep^L_0 (\beta )  = \frac{mR}{2} e^{-\beta } , ~~~~~~ 
\vep^R_0  (\beta ) = \frac{mR}{2} e^{\beta} 
\eeq
The structure of the TBA is the following:
\barray
\vep^L &=& \vep^L_0  - G *  L^R  
 \nonumber
\\
\label{masslessTBA.2}
\vep^R &=& \vep^R_0  - G * L^L  
\earray
where 
\beq
\label{masslessTBA.3}
L^{L,R} = \log \( 1 + \exp ({-\vep^{L,R}}) \) 
\eeq
The effective central charge is:
\beq
\label{masslessTBA.4}
\ceff (mR) = \frac{6}{\pi^2} \real 
\int_{-\infty}^\infty  d\beta ~ \( \vep^R_0  L^R + \vep^L_0  L^L \) 
\eeq
When the left-right pseudo-energies are decoupled,  then 
by simple shifts of the rapidity one can show
that $\ceff (e^l R) = \ceff (R)$ for {\rm any}  $l$ indicating
a scale invariant conformal field theory with $\ceff$ independent of
$R$. 
When there is non-trivial LR scattering, which leads to the terms in
eq. (\ref{masslessTBA.2}) which couple L and R, then $\ceff$ becomes
a function of $mR$. 

Now consider an S-matrix that is periodic in rapidity, as for
our model, such that $G(\beta + \lambda) = G(\beta )$,  where
as usual $\lambda$ is the period of the RG.  
  An example of such an S-matrix for a single scalar
particle was given in  \cite{Mussardo}.  
The due to the periodicity  of the kernels $G$,
$\ceff$ is now  invariant under a discrete scale transformation
$e^\lambda$. 
To see this,  one notes, using the periodicity of the kernels, 
 that $\vep_L (\beta - \lambda)$ and
$\vep_R (\beta + \lambda)$ satisfy the same integral equations 
as $\vep_{L,R}$ but with $R$ replaced by $e^\lambda R$. 
Namely,
\beq
\label{masslessTBA.5}
\vep^L (\beta - \lambda, R) = \vep^L (\beta, e^\lambda R), 
~~~~~\vep^R (\beta + \lambda, R) = \vep^R (\beta , e^\lambda R)
\eeq
Using these equations in eq. (\ref{masslessTBA.4}) one can verify
the periodicity of $\ceff$:
\beq
\label{masslessTBA.6}
\ceff( e^\lambda R ) = \ceff (R)
\eeq

The above exact periodicity of $\ceff$ over all length scales
is rather remarkable and entirely consistent 
 with the RG argument that led to 
eq. (\ref{cRG.3}).  Also, it indicates that unlike the massive
case,  there is  no scale dependence of the amplitudes of oscillation
of $\ceff$ and thus 
no singularities of the kind encountered in the last section.

\newpage 

\section{Conclusions}

We have analyzed the finite size effects, equivalently the finite
temperature effects,  in the ground state energy
or effective central charge $\ceff$ for  the massive
S-matrix theory proposed in  \cite{LRS}.  
The  massive S-matrix was conjectured to describe the cyclic
regime of the Kosterlitz-Thouless flows \cite{LRS}.   
The S-matrices are periodic in the rapidities,  with a period
consistent with the RG analysis.   This periodicity leads to 
novel finite size effects which also have periodic structures
consistent with the cyclic RG. 
 We obtained an approximate analytic expression for
$\ceff$ expressed in terms of Riemann's zeta function 
eq. (\ref{clinear}), which has log-periodic behavior and
agrees remarkably well with the exact numerical solution.

For the massive case,  there is a wide range of scales in the UV  where
$\ceff$ oscillates.  However in the very deep UV a singularity 
is encountered,  and in some regions  $\ceff$ is not well defined
(coexistence regions described in section V).   This suggests 
the massive S-matrix theory needs an explicit UV cutoff in the 
very deep UV.   For the massive case $\ceff$ also lacks a periodicity
on all scales as would be predicted by the RG.

For the massless case the TBA equations are unknown and we could
not carry out the comparable analysis.  However we argued on
general grounds that  
the massless case should have some appealing properties. 
The $\ceff$ should be exactly  periodic on all scales, consistent
with the RG.  We hope to report on this in a future publication.

The log-periodic dependence of $\ceff(mR)$ on $mR$ 
is reminiscent of the log-periodic fractal dimensions
of self-similar fractals, such as the Cantor's triadic set. 
Self-similarity amounts to discrete
scale invariance, which in turn leads to log-periodic
behavior in a wide variety of problems such as
DLA (diffusion-limited-aggregation clusters), 
earthquakes, financial crashes, etc.
 \cite{sornette,log,tierz}.

On a broader note, there
appears  to be a network of deeply interrelated  
concepts and techniques, namely  
RG limit cycles, discrete scale invariance, complex exponents, 
fractals, log-periodicity, quantum groups (with real $q$), 
zeta function regularizations, number theory, etc, whose
full  significance needs to be clarified.

\section*{Acknowledgments} 

We would like to thank D. Bernard, A. Capelli, 
J. Cardy, D. Friedan, I. Klebanov,  S. Lukyanov,
G. Moore,  G. Mussardo, and H. Saleur for discussions.
JMR also thanks the warm welcome received from 
people at the Department of Theoretical 
Physics in the University of Zaragoza. 
This work has been supported by the Spanish grants 
BFM2000-1320-C02-01 (JMR and GS), and by the NSF of the USA. 
We also thank the EC Commission for financial support
via the FP5 Grant HPRN-CT-2002-00325.

\section*{Appendix A}

\def\K{K^{(a)}}
\def\bK{\bar{K}^{(a)}}
\def\hK{\hat{K}^{(a)}}
\def\hbK{\hat{\bar{K}}^{(a)}}
\def\bs{\bar{\sigma}}
\def\Kbar{K'} 

In this appendix we compute the 
coefficients  $F_n= \{ L_{0,n}, K_n, \Kbar_n \}$
of the functions $ F(\rap) = \{
L_0(\rap) , K(\rap) , \Kbar(\rap) \} $ 
needed in section V-D.   They  are given by the formula
\beq
F_n = \frac{h}{2 \pi} \hat{F} (nh) 
\label{ap3}
\eeq
where
$\hat{F} (t) $ is the Fourier transform of $F$:
\beq
\hat{F}(t) \equiv \int_{-\infty}^\infty 
d \rap  \; F(\rap ) \; e^{- i t \rap }
\label{ap2}
\eeq

\subsection{Fourier transform of $K(\rap )$}

The function $K(\rap )$ is given by 
\beq
K(\rap )   = \frac{1}{1 + \exp(r \cosh(\rap ))} 
\label{ap4}
\eeq
where $r = m R$ is the scaling variable. 
Since $K(\rap )$ is an even function we can write
its Fourier transform (\ref{ap2}) as
\beq
\hat{K}(t) \equiv \int_{-\infty}^\infty 
d \rap  \; \frac{ e^{ i t \rap }}{ 1 + \exp(r \cosh(\rap ))}
\label{ap5}
\eeq

For $t >0$ one can evaluate (\ref{ap5}) 
as the sum of the residues
of the poles of $K(\rap )$ in the upper half plane ${\imag } \rap  > 0$.
These poles are the zeros of the denominator of $K(\rap )$
and  are given by
\barray
\rap ^{\pm}_{p, p'} & = & \pm \theta_p +
i \frac{\pi}{2} (1 + 2 p'), \quad p, p' \geq 0  \label{ap6} \\
\theta_p & = & \sinh^{-1} \left( \frac{ \pi ( 1 + 2p)}{r}
\right)  
\nonumber 
\earray
The residue formula yields
\beq
\hat{K}(t) =
- \frac{ 2 \pi}{r \cosh(\pi t/2)} \; 
\sum_{p=0}^\infty \frac{ \cos( t \theta_p)}{\cosh \theta_p}
\label{ap7}
\eeq
For $t > 0$ and real this series diverges. We shall regularize it
using the Riemann zeta function $\zeta(z)$. For simplicity let us
consider the UV limit $r \rightarrow 0$ where $\theta_p$ has the expansion,
\beq
\theta_p = s
+ \log( 1 + 2p) + \frac{e^{-2 s}}{( 1 + 2p)^2} + O(e^{-4 s})
\label{ap8}
\eeq
where $s = \log(2 \pi/r)$.  
The series (\ref{ap7}) becomes
\beq
\hat{K}(t) =
- \frac{1}{\cosh(\pi t/2)} \; 
\sum_{p=0}^\infty 
\left[ e^{i t s} \left( 
\frac{1}{(1 + 2p)^{1 - it}} - e^{- 2s} 
\frac{(2 - it)}{( 1 + 2p)^{3 - it}} \right) + (t \rightarrow -t) \right]
+ O(e^{- 4 s}) 
\label{ap9}
\eeq

The sums in $p$ can be done using the Riemann zeta function
$\zeta(z)$ 
\beq
\zeta(z) = \sum_{n=1}^\infty \; \frac{1}{n^z} 
\label{ap10}
\eeq
together with the relation:
\beq
\sum_{p=0}^\infty \frac{1}{( 1 + 2p)^z} =
(1 - 2^{-z}) \;  \zeta(z) \equiv \zeta_1(z) 
\label{ap11}
\eeq
This gives 
\beq
\hat{K}(t) =
- \frac{1}{\cosh(\pi t/2)} \; 
\left[ e^{i t s} 
\left[ \zeta_1(1 - i t) 
- e^{- 2 s} (2 - i t) \zeta_1(3 - it) 
\right] + (t \rightarrow  -t) \right] + O(e^{- 4 s}) 
\label{ap12}
\eeq

Strictly speaking eqs.(\ref{ap10}) and (\ref{ap11})  
are only defined when ${\real } \; z > 1$. However the Riemann
zeta function is defined in the entire complex plane
except at $z=1$ where it has a pole. 
Our use of $\zeta(z)$ is completely analogous to other situations 
in quantum field theory  where it serves to regularize
formal expressions.

Equation (\ref{ap3}) gives the Fourier coefficients
\beq
K_n = \frac{h}{2 \pi} \; \hat{K}(n h)
\label{ap13}
\eeq
In the UV limit the terms proportional to $e^{-2 p s} \;\; (p=1,2, \dots)$  
vanish very rapidly and one is left with the oscillating terms
that depend on $\zeta_1(1  \pm i t)$. 
The value $K_0$ is special since $\zeta(z)$ behaves near $z=1$
as
\beq
\zeta(z) \sim \frac{1}{z-1} + \gamma 
\label{ap14}
\eeq
where $\gamma$ is the Euler's constant. Using this equation  the value
of $K_0$ is given by
\beq
K_0 = \frac{h}{2 \pi} \left( s - \gamma - \log 2 + \frac{7 \zeta(3)}{2}
e^{- 2 s} \right)  + O(e^{-4 s})
\label{ap15}
\eeq

The linear dependence of $K_0$ with $s$ in the UV limit
can be easily understood using the ``kink method'' proposed
by Zamolodchikov to study the UV limit of usual massive models  \cite{ZamoTBA}.
From eq.(\ref{ap2},\ref{ap3})  we have 
\beq
K_0 = \frac{h}{2 \pi} \int_{- \infty}^\infty d \rap 
\; \frac{1}{1 + \exp ({r \cosh \rap})}
\label{ap16}
\eeq 
which in the limit $r \ll 1$ can be approximated as
\beq
K_0 = \frac{h}{\pi} \int_{0}^\infty d \rap 
\; \frac{1}{1 + \exp(e^{\rap - s'}) }
\label{ap17}
\eeq
where $s' = \log(2/r) \gg 1$. 
From this expression it is clear that the integrand
remains close to $1/2$ for $0 < \rap \lesssim s'$ and that it vanishes
exponentially for $ \rap  \gg s'$. Hence $K_0 \sim h s'/(2 \pi)$ 
which agrees with eq.(\ref{ap15}) up to constant terms
since $s' = s - \log \pi$.

\subsection{Fourier transform of $\Kbar$}

The function $\Kbar(\rap )$ is defined as
\beq
\Kbar(\rap)  = \vep_0(\rap ) K(\rap )
\label{ap34}
\eeq
Following the same steps as in the previous subsection we find
the Fourier transform,
\beq
\label{Kbarhat}
\hat{\Kbar} (t) = 
\frac{ 2 \pi}{ \sinh(\pi t/2)} \; 
\sum_{p=0}^\infty \tanh(\theta_p) \sin( t \theta_p)
\eeq
In the UV limit this can be approximated as 
\beq 
\hat{\Kbar}(t) =
- \frac{i \pi}{\sinh(\pi t/2)} \; 
\left[ e^{i t s} 
\left[ \zeta_1(- i t) 
- e^{- 2 s} (2 - i t) \zeta_1(2 - it) 
\right] - (t \rightarrow  -t) \right] + O(e^{- 4 s}) 
\label{ap22}
\eeq
where we have included explicitly the term of order $e^{- 2s}$
as in eq.(\ref{ap12}). The value of this function at $t=0$
is given by 
\beq
\hat{\Kbar}(0) = - 4 \zeta(0) \, \log 2 = 2 \log 2 
\label{ap41}
\eeq
where we have used that $\zeta(0) = -1/2$. 

Using the kink method
one can see that this result is equivalent to the following integral
\beq
2 \log 2 = \frac{1}{2} \int_{- \infty}^\infty d\rap ~  
\frac{e^\rap }{ 1 + \exp( e^\rap /4 ) }
\label{ap42}
\eeq
which can be checked numerically to any precision.

\subsection{Fourier transform of $L_0(\rap )$}

The function $L_0(\rap )$ is defined as
\beq
L_0(\rap ) = \log( 1 + e^{- r \cosh \rap })
\label{ap43}
\eeq
To compute its Fourier transform  we note  that
\beq
- r \frac{d}{dr} L_0(\rap )  = \frac{d}{ds} L_0 (\rap ) 
=  \Kbar (\rap ) 
\label{ap44}
\eeq
Hence the Fourier transform of $L_0(\rap )$ can be obtained
integrating eq.(\ref{ap44})  in $s$, obtaining:
\beq
\hat{L_0}(t) =
- \frac{\pi}{\sinh(\pi t/2)} \; 
\left[ e^{i t s} 
\left( 
\frac{1}{t} \zeta_1(- i t) 
+ i e^{- 2 s} \zeta_1(2 - it) 
\right) 
+ 
 e^{- i t s} 
\left(
\frac{1}{t} \zeta_1(i t) 
- i e^{- 2 s} \zeta_1(2 + it) 
\right) 
\right]   
\label{ap45}
\eeq

The integration constant is zero as we shall next show. 
Let us consider the value of $\hat{L_0}(0)$: 
\beq
\hat{L_0}(0) = 
2 \, s \log 2  - \log 2 \,  \log (2 \pi^2)
\label{ap46}
\eeq
where we have used that $\zeta'(0) = - \frac{1}{2} \log (2 \pi).$
Notice that the leading term in $s$ coincides with the value
of $\hat{\Kbar}(0)$ given in eq.(\ref{ap41}), 
but there is a constant term that
appears in the limit $t \rightarrow 0$
in (\ref{ap45}). Using the kink method 
eq. (\ref{ap45}) amounts to, 
\beq
\lim_{r \rightarrow 0} \left[ 2 
\int_0^\infty d\rap  \log( 1+ \exp( - r \cosh \rap )) - 2 s \, \log2  \right]
= - \log 2 \,  \log (2 \pi^2),\quad\quad  s = \log(2 \pi/r)
\label{ap47}
\eeq 
which we  checked numerically.  This shows that 
the integration constant relating  $\hat{\Kbar }(t)$
and  $\hat{L_0}(t)$ vanishes.

\section*{Appendix B}

In this appendix we justify the analytic approximation made
in section V-D of the non linear eq.(\ref{4.16}). 

The full expansion of the function $L(\rap)$ is given by
\beq
L(\rap) = L_0(\rap) - \sum_{a=1}^\infty
\frac{\s^a}{a!}\;  K^{(a-1)}(\rap)
\label{ap-b1}
\eeq
\no where  
\beq
 K^{(a)}(\rap) = \frac{d^a K}{d \vep_0^a}, \qquad
K^{(0)} \equiv K =  \frac{1}{1 + e^{\vep_0}}
\label{ap-b2}
\eeq
\no Using eq.(\ref{ap-b1}), the expansion of eq.(\ref{4.16}) to all orders
in $\s$ becomes, 
\beq
\label{ap-b3}
\sum_m  A_{n,m} \sigma_m = B_n + \sum_{a=2}^\infty 
\frac{1}{a!} \sum_{m_1, \cdots, m_a } 
C^a_{n, \; \sum m_i} \; \;  \s_{m_1} \cdots \s_{m_a} 
\eeq
\no where $A_{n,m}$ and $B_n$ were defined in eq.(\ref{4.23}) 
and 
\barray
\label{ap-b4} 
C^a_{n,m} & = &  g_{0,n} K^{(a-1)}_{n-m}  
- g_{1,n} K^{(a-1)}_{-n-m},  \\
K^{(a)}_n &= & \frac{h}{2 \pi} \int_{- \infty}^\infty
d \rap \; e^{- i n h \rap} K^{(a)}(\rap) = \frac{h}{2 \pi} 
\hat{K}^{(a)}(n h) 
\nonumber 
\earray
The analytic approximation of section V-D consists in i) eliminating 
the terms O($\s^n$), with ${n \ge 2} $,  
in eq.(\ref{ap-b3}) yielding eq.(\ref{4.23}),   
and ii) considering 
only the coefficient $K_0$ which depends linearly on 
the scale $s$ (recall eq.(\ref{4.27})). The latter asumption
is justified if the $\s_n$'s are small enough. As for the first
assumption we shall show below that the terms in eq.(\ref{ap-b3})
containing high powers in $\s_n$ do not depend linearly 
on the scale $s$, and can thus be neglected in the UV. 

The Fourier transform of $K^{(a)}$ can be found 
by the techniques of appendix A and it reads, 

\beq
\hK(t) = -  \left( \frac{ - i }{\pi} \right)^a
\sum_{\nu= \pm 1} e^{ i t \nu s} \; 
\frac{ \Gamma( a + 1 - i t \nu)}{ \Gamma( 1 - i t \nu)}
\; \zeta_1( a +1 - i t \nu) \times
\left\{
\begin{array}{ll}
1/\cosh(\pi t/2), & a =0, 2,\dots \\
\nu/ \sinh(\pi t/2), & a= 1, 3, \dots 
\end{array}
\right.
\label{ap-b5}
\eeq

\no The case $a=0$ agrees with eq. (\ref{ap12}). Let us focus
on the cases where $a \ge 1$. 
Only for $t=0$ one may get a linear dependence on $s$, namely
\beq
\hK(0) = a! \; \zeta_1( a +1) \times \left\{
\begin{array}{ll}
- 2 \left( \frac{i}{\pi} \right)^a, &  a =2, 4, \dots \\
 4 \left( \frac{i}{\pi} \right)^{a+1} s, &  a =1, 3, \dots \\
\end{array}
\right.
\label{ap-b6}
\eeq
\no Returning to eq.(\ref{ap-b3}),  
the only terms that may contribute linearly in $s$, are
\beq
\label{ap-b7}
\sum_{a \ge 2, \;  {\rm even}}
s \; \frac{2 h \; \zeta_1(a) i^a}{a \; \pi^{a+1}}  \left(
g_{0,n} \sum_{{m_1}, \dots, {m_a}} 
\delta_{n, m_1 + \dots + m_a} \;  \s_{m_1} \dots \s_{m_a}
-
g_{1,n} \sum_{{m_1}, \dots, {m_a}} 
\delta_{-n, m_1 + \dots + m_a} \;  \s_{m_1} \dots \s_{m_a}
\right)  
\eeq
\no Using eq.(\ref{4.18}) one can show that
\beq
\label{ap-b8}
\sum_{{m_1}, \dots, {m_a}} 
\delta_{n, m_1 + \dots + m_a} \;  \s_{m_1} \dots \s_{m_a}
= (-1)^a e^{- n \pi h} 
 \sum_{{m_1}, \dots, {m_a}} 
\delta_{-n, m_1 + \dots + m_a} \;  \s_{m_1} \dots \s_{m_a} 
\eeq
\no which together with eq.(\ref{4.14}), namely 
$g_{1,n} = e^{-n \pi h} g_{0,n}$, 
implies the vanishing of (\ref{ap-b7}). In summary, all the linear
dependence in $s$ of the eq.(\ref{ap-b3}) comes from the coefficient
$B_n$.

\end{document}